\documentclass[fleqn,usenatbib]{mnras}

\usepackage{newtxtext,newtxmath}

\usepackage[T1]{fontenc}
\usepackage{ae,aecompl}

\usepackage{graphicx}	
\usepackage{amsmath}	
\usepackage{amssymb}	

\title[Probing Temperature Structure Using Polarization]{Probing the Temperature Structure of Optically Thick Disks Using Polarized Emission of Aligned Grains}

\author[Z. D. Lin et al.]{
Zhe-Yu Daniel Lin$^{1}$\thanks{E-mail: zdl3gk@virginia.edu}, 
Zhi-Yun Li$^{1}$, 
Haifeng Yang$^{2}$, 
Leslie Looney$^{3}$, 
\newauthor
Chin-Fei Lee$^{4,5}$, 
Ian Stephens$^{6}$, 
and Shih-Ping Lai $^{7}$
\\
$^{1}$Department of Astronomy, University of Virginia, Charllottesville, VA 22904, USA \\
$^{2}$Institute for Advanced Study, Tsinghua University, Beijing, 100084, People's Republic of China \\
$^{3}$Department of Astronomy, University of Illinois at Urbana-Champaign, Urbana IL 61801, USA \\
$^{4}$Academia Sinica Institute of Astronomy and Astrophysics, P.O. Box 23-141, Taipei 106, Taiwan \\
$^{5}$Graduate Institute of Astronomy and Astrophysics, National Taiwan University, No. 1, Sec. 4, Roosevelt Rd., Taipei 10617, Taiwan \\
$^{6}$Harvard-Smithsonian Center for Astrophysics, 60 Garden Street, Cambridge, MA 02138, USA \\
$^{7}$Institute of Astronomy and Department of Physics, National Tsing Hua University, Hsinchu 30013, Taiwan
}

\date{Accepted XXX. Received YYY; in original form ZZZ}

\pubyear{2019}

\begin{document}
\label{firstpage}
\pagerange{\pageref{firstpage}--\pageref{lastpage}}
\maketitle

\begin{abstract}
Polarized continuum emission from aligned grains in disks around young stellar objects can be used to probe the magnetic field, radiation anisotropy, or drift between dust and gas, depending on whether the non-spherical grains are aligned magnetically, radiatively or mechanically. We show that it can also be used to probe another key disk property -- the temperature gradient --  along sight lines that are optically thick, independent of the grain alignment mechanism. We first illustrate the technique analytically using a simple 1D slab model, which yields an approximate formula that relates the polarization fraction to the temperature gradient with respect to the optical depth $\tau$ at the $\tau=1$ surface. The formula is then validated using models of stellar irradiated disks with and without accretion heating. The promises and challenges of the technique are illustrated with a number of Class 0 and I disks with ALMA dust polarization data, including NGC 1333 IRAS4A1, IRAS 16293B, BHB 07-11, L1527, HH 212 and HH 111. We find, in particular, that the sight lines passing through the near-side of a highly inclined disk trace different temperature gradient directions than those through the far-side, which can lead to a polarization orientation on the near-side that is orthogonal to that on the far-side, and that the HH 111 disk may be such a case. Our technique for probing the disk temperature gradient through dust polarization can complement other methods, particularly those using molecular lines.  

\end{abstract}



\section{Introduction}
Dust polarization observations are traditionally used to trace the magnetic field, based on the idea that non-spherical grains preferentially align their long axis perpendicular to the field \citep[e.g.][]{Draine1997, Hildebrand2000, Lazarian2007, Andersson2015}. At optical wavelengths, intrinsically unpolarized star light are preferentially absorbed along the long axis of foreground grains and produce polarization parallel to the magnetic field. Grain thermal emission, on the other hand, preferentially emit light with polarization along the long axis and produce polarization perpendicular to the magnetic field. This dichroism allows measurements of the magnetic field morphology over a wide range of scales, from molecular clouds to protostellar envelopes \citep[e.g.][]{PlanckCollaboration2016, Girart2006, Stephens2013}. It is widely believed that magnetic fields play a key role in determining protoplanetary disk structure and evolution through magnetorotational instability \citep[e.g.][]{Balbus1991} and magnetocentrifugal disk wind \citep[e.g.][]{Blandford1982}, which set the conditions of planet formation \citep{Morbidelli2016}.

Recently, polarized (sub)millimeter emission has been detected in an increasing number of disks by Atacama Large Millimeter/submillimeter Array (ALMA) with its high sensitivity and angular resolution. However, the origin of disk polarization remains uncertain, since grains do not have to be aligned with just the magnetic field \citep{Kataoka2017, Yang2019}. They may also be aligned in the direction of the radiative anisotropy \citep{Lazarian2007, Tazaki2017} or the drift velocity of the grains relative to the ambient gas \citep{Gold1952, Lazarian1995,Lazarian2007_MAT}.
Furthermore, even spherical grains can produce polarized emission by self-scattering of large grains in an anisotropic radiation field \citep{Kataoka2015, Yang2016_HLTau, Yang2017_nearfar, Stephens2019_whitepaper}. The scattering interpretation of the disk polarization is favored in several targets \citep[e.g.,][]{Bacciotti2018, Dent2019, Hull2018, Kataoka2016, LeeCF2018, Stephens2014,Stephens2017, Girart2018, Harris2018}. One way to gauge the effects of scattering and identify polarization from aligned grains would be to observe at multiple wavelengths since the efficiency for scattering for grains of given sizes decreases rapidly with the wavelength in the optically thin and small-particle (or Rayleigh scattering) limit. Indeed, in the disk of Class I protostar BHB 07-11, \cite{Alves2018} detected polarization with ALMA at three wavebands (Bands 3, 6, and 7 or $\sim$ 3mm, 1.3mm and 0.87 mm, respectively) with consistent polarization orientations across three bands and increasing polarization fraction with wavelength, which is generally not expected for scattering-induced polarization. The rather high mean polarization fractions ($\sim$ 7.9, 5.3 and 3.5$\%$ for Bands 3, 6, and 7 respectively) are also higher than those typically produced in models of scattering-induced disk polarization ($\sim 1 \%$). At least for this well-studied source, scattering is unlikely the main mechanism for producing the observed multi-wavelength disk polarization and aligned grains are favored.   

In this paper, we aim to improve interpretations of disk polarization from aligned grains by including effects of temperature gradient and optical depth. The analysis for polarization by aligned grains has traditionally been carried out under the assumption of isothermal grains \citep{Hildebrand2000}. However, polarization from aligned grains can be changed both qualitatively and quantitatively by a temperature gradient along the line of sight \citep{Yang2017_nearfar, Liu2018}. This is particularly relevant to protoplanetary disks, which are expected to be far from isothermal with strong temperature gradients in both radial and vertical directions \citep[e.g.][]{DAlessio1998,Dullemond2002_varied, Dullemond2002_2D}. Although the detailed temperature structure for disks remain poorly constrained empirically, observations have started to show evidence for spatially varying disk temperatures \citep[e.g.,][]{Rosenfeld2013, Cleeves2016, Pinte2018}. We are thus motivated to explore how temperature gradient affects polarization due to aligned grains with the goal of developing a method to probe disk temperature gradient using polarization.

The structure of this paper is as follows: we start with an analytical model to illustrate the effects of temperature gradient and optical depth in Section \ref{sec:analytical}. In Section \ref{sec:ObsModeling}, we provide a disk model and compute the temperature using the Monte Carlo radiative transfer code RADMC-3D\footnote{\url{http://www.ita.uni-heidelberg.de/~dullemond/software/radmc-3d/}}. The model is then used to create the expected polarization images at (sub)millimeter wavelength. We demonstrate how polarization can be used as a direct probe for temperature gradient. The results are discussed in the context of ALMA dust polarization observations in Section \ref{sec:Discussion}. We summarize the main results in Section \ref{sec:Conclusions}. 

\section{Analytical Illustration} \label{sec:analytical}

\begin{figure}
    \centering
    \includegraphics[width=\columnwidth]{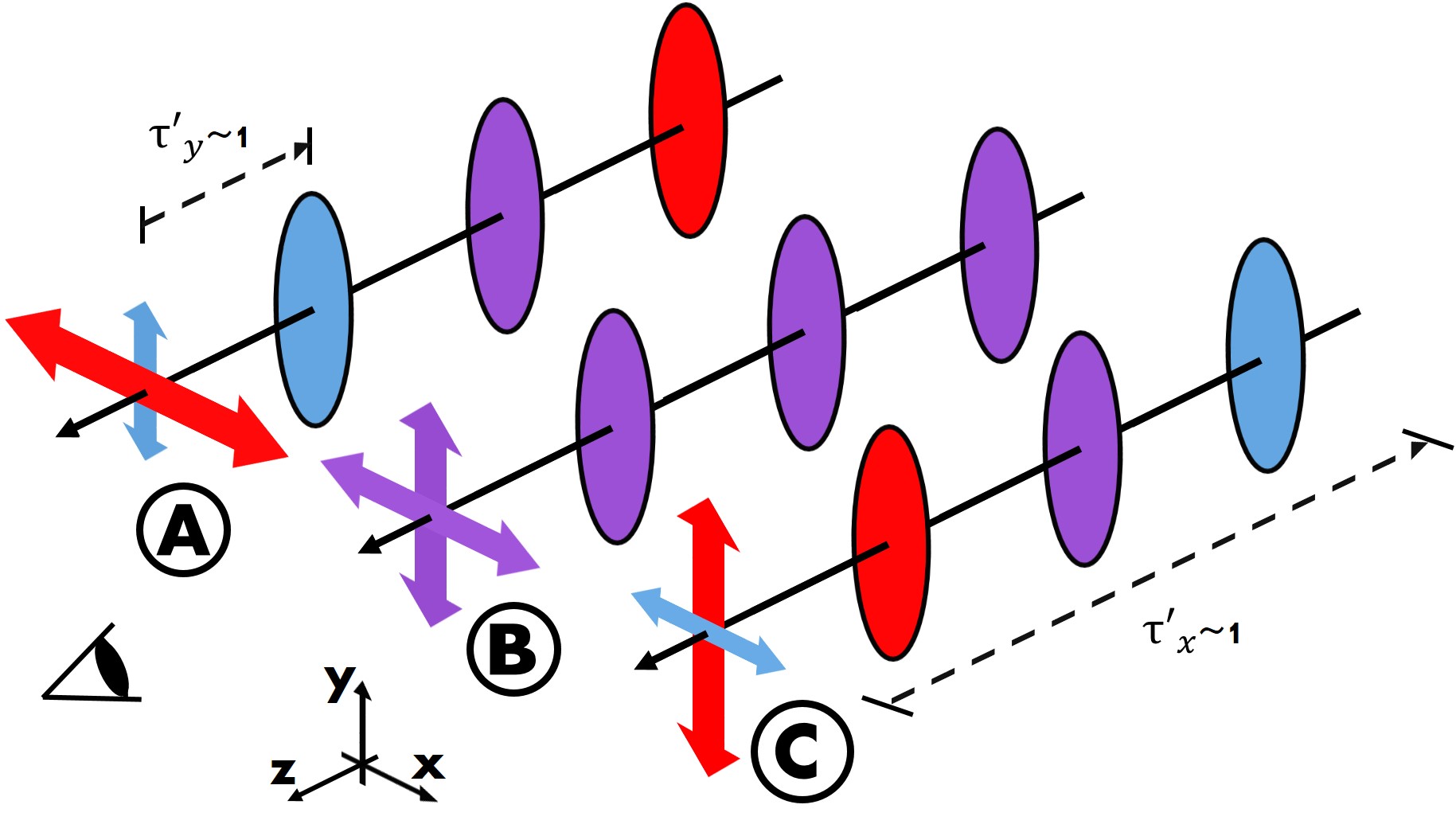}
    \caption{
    Schematic figure depicting observed polarization direction with different temperature gradients in the optically thick limit. The definition of the $x-$ and $y-$directions are shown in the lower-left of the figure. The observer is located in the positive $z-$direction. The oblate grains have their minor axes aligned in the $x-$direction and are threaded by black arrows which represent the direction light travels to the observer. The colored ellipses represent the projected cross sections to the observer of such oblate grains, which are longer in the $y-$direction and shorter in the $x-$direction. Red, purple, blue colors represent hot, moderate, and cold grains. The dashed arrows represent the location where the optical depth in the $x-$ or $y-$direction becomes unity. The size of the double-sided arrows represents the amount of emission for the light polarized in either the $x-$ or $y-$direction. The color of the arrows corresponds to the temperature of the grain mainly responsible for that emission, i.e. the location where optical depth is unity. Since intensity represented by the red arrows originating from hotter grains are larger than that from colder grains, the observed polarization direction follows the red arrow. For Case B (isothermal), the emission from the $x-$ and $y-$directions are equal and thus no polarization is observed. For Case A, the temperature increases along the line of sight, away from the observer, and the polarization direction is in the $x-$direction and vice versa for Case C. 
    }
    \label{fig:schem}
\end{figure}

In this section we use a simple model to illustrate the temperature gradient effects on polarization from aligned grains. We assume a semi-infinite constant density slab, with $x-$ and $y-$directions extending to infinity, but with a finite thickness, $\Delta z$, in the $z-$direction (see Figure \ref{fig:schem}). The observer is in the positive $z-$direction thus the line of sight is (and optical depth increases) in the negative $z-$direction. The signed polarization fraction is defined by $p \equiv (I_{x} - I_{y}) / (I_{x} + I_{y})$ where $I_{x}$ and $I_{y}$ are the specific intensity at the surface of the slab facing the observer along the direction perpendicular to the slab. The sign of polarization determines the orientation of polarization. Positive polarization means that the polarization is parallel to the $x-$direction in our set up. Negative polarization means the polarization is in the $y-$direction.

Since aligned grains generally spin rapidly around their shortest axes and act effectively as oblate grains independent of their shapes when ensemble-averaged, we consider oblate grains\footnote{We will discuss in Section~\ref{sec:EdgeOn} below effectively prolate grains, which are appropriate for aerodynamically aligned grains \citep{Gold1952}.} that have their short axes aligned in the $x-$direction. With this orientation, the oblate grains are viewed edge-on and the opacities in the $x-$ and $y-$directions, or $\kappa_{x}$ and $\kappa_{y}$, are not equal, but related by $\kappa_{x} < \kappa_{y}$. The ``intrinsic polarization'' is defined as 
\begin{equation}
    p_{0} \equiv (\kappa_{x} - \kappa_{y}) / (\kappa_{x} + \kappa_{y}) \text{ ,}
\end{equation}
which is the polarization of these edge-on oblate grains in the optically thin limit as we will see later. 

In the absence of any background source, the solutions to the radiative transfer equation for the $x-$ and $y-$polarized light are
\begin{align} 
    I_{x} &= \int_{0}^{\tau_{x}} \dfrac{1}{2} S(\tau') e^{-\tau'_{x}}d\tau'_{x} \label{eq:Ix} \\
    I_{y} &= \int_{0}^{\tau_{y}} \dfrac{1}{2} S(\tau') e^{-\tau'_{y}}d\tau'_{y} \label{eq:Iy} \text{ ,}
\end{align}
where $\tau_{x}$ and $\tau_{y}$ are the total optical depths of the slab in the corresponding directions, $\tau'_{x}$ and $\tau'_{y}$ are the variable optical depths, and $S$ is the source function. The mean variable optical depth, $\tau'$, is given by $\tau' \equiv (\tau'_{x}+\tau'_{y})/2$. The total optical depth in the $x-$direction is related to the opacity by $\tau_{x} = \kappa_{x} \rho \Delta z$, where $\rho$ is the density of the slab, and likewise for that in the $y-$direction. Since we are ignoring scattering, the source function is determined by the temperature related through the black body radiation. To consider temperature variations along the line of sight, we let the source function for the total intensity be linear to optical depth, 
\begin{equation} \label{eq:deltaS}
    S(\tau') = S_{0} + \Delta S \tau' \text{, where } \Delta S \equiv \dfrac{d S}{d \tau'} \text{ ,}
\end{equation} where $S_{0}$ is the source function at the surface of the slab. We also define the mean total optical depth as $\tau \equiv (\tau_{x} + \tau_{y})/2$. Note that since optical depth increases into the slab, a positive source function gradient, $\Delta S > 0$, means the source function increases along the line of sight, and vice versa. Solving the emitted intensity (Equations (\ref{eq:Ix}) and (\ref{eq:Iy})) using the source function (Equation (\ref{eq:deltaS})), one obtains
\begin{align} \label{eq:Ixytau}
    I_{x} &= \dfrac{1}{2} S_{0} \big[ 1 - e^{-\tau_{x}} \big] + \dfrac{1}{2} \dfrac{\Delta S}{1+p_{0}} \big[ 1 - (1 + \tau_{x}) e^{-\tau_{x}} \big] \\
    I_{y} &= \dfrac{1}{2} S_{0} \big[ 1 - e^{-\tau_{y}} \big] + \dfrac{1}{2} \dfrac{\Delta S}{1-p_{0}} \big[ 1 - (1 + \tau_{y}) e^{-\tau_{y}} \big] \text{ .}
\end{align}
With $I_{x}$ and $I_{y}$ solved, we will consider the produced polarization in various limits. 

By considering $\Delta S=0$, one recovers the polarization in the isothermal case \citep{Hildebrand2000}:
\begin{equation} \label{eq:poliso}
    p = \dfrac{
            e^{-\tau}\sinh(p_{0}\tau)
                }{
            1 - e^{-\tau}\cosh(p_{0}\tau)
                } \text{ .}
\end{equation}
In the optically thin limit for the isothermal case, $\tau \ll 1$, the polarization is $p=p_{0}$, thus the polarization measured for grains viewed edge-on in the optically thin limit is the intrinsic polarization. Since $\kappa_{x}<\kappa_{y}$, the intrinsic polarization in our set up is negative and thus the intrinsic polarization is along the $y-$direction. Note that in the optically thick limit, $\tau \gg 1$, polarization in this isothermal case is zero. 

For the non-isothermal case, $\Delta S \ne 0$, and in the optically thin limit, 
\begin{equation} \label{eq:pthin0}
    p = \dfrac{
            S_{0}(\tau_{x}-\tau_{y}) + \Delta S \bigg(\dfrac{1}{1+p_{0}}\tau_{x}^{2} - \dfrac{1}{1-p_{0}}\tau_{y}^{2} \bigg) 
                }{
            S_{0}(\tau_{x}+\tau_{y}) + \Delta S \bigg(\dfrac{1}{1+p_{0}}\tau_{x}^{2} + \dfrac{1}{1-p_{0}}\tau_{y}^{2} \bigg) 
                } \text{ .}
\end{equation}
It is clear that non-isothermal effects are second order in optical depth and does not strongly affect the polarization when the slab is optically thin. If the $\tau^{2}$ terms are ignored when $\tau \ll 1$, polarization becomes $p = (\kappa_{x} - \kappa_{y}) / (\kappa_{x} + \kappa_{y})$, or the intrinsic polarization, $p_{0}$. 

In the optically thick limit, polarization approaches 
\begin{equation} \label{eq:poldSthick}
    p = \dfrac{- \Delta S p_{0}}{S_{0}(1-p_{0}^{2}) + \Delta S} \text{ .}
\end{equation}
Thus, in contrast to the isothermal optically thick case (Equation (\ref{eq:poliso})), there is a non-zero level of polarization and also an extra negative sign. If $\Delta S < 0$, the observed polarization will remain in the same direction as $p_{0}$. However, when $\Delta S >0$, the observed polarization has an opposite sign relative to $p_{0}$, thus the polarization direction is perpendicular to the intrinsic polarization. We will term this change in sign as ``polarization reversal."

In the Rayleigh-Jeans limit, assuming a black body radiation, the source function scales linearly with temperature, thus Equation (\ref{eq:deltaS}) becomes $T = T_{0} + \Delta T \tau'$ with $\Delta T \equiv d T / d \tau'$ and $T_{0}$ being the temperature at the surface of the slab. Using Equation (\ref{eq:poldSthick}), polarization is related to temperature by 
\begin{equation} \label{eq:poldTthick}
    p = \dfrac{- \Delta T p_{0}}{T_{0}(1-p_{0}^{2}) + \Delta T} \text{ .}
\end{equation}
Note Equation (\ref{eq:poldTthick}) is basically Equation (10) in \cite{Yang2017_nearfar}. Furthermore, in the limit where $p_{0}^{2} \ll 1$,  
\begin{equation} \label{eq:polpsi}
    - \dfrac{p}{p_0} \approx \dfrac{\Delta T} {T_{1}} = \dfrac{d \ln{T} }{d \tau'} \bigg\vert_{\tau'=1}\equiv \psi 
\end{equation}
where $T_{1}$ is the temperature at $\tau'=1$. Therefore, the negative of the ratio of polarization to intrinsic polarization is simply the gradient of the logarithmic temperature, $d \ln T / d \tau'$, at the $\tau'=1$ surface, which we denote by a dimensionless parameter $\psi$. Since the observed disk polarization is typically found to be a few percent, we will be using Equation (\ref{eq:polpsi}) in Section \ref{sec:ObsModeling} below. Note that Equation (\ref{eq:polpsi}) is essentially the Eddington-Barbier relation for polarization in the Rayleigh-Jeans limit: while the observed intensity reflects the temperature at the $\tau'=1$ surface, the polarization measures the logarithmic temperature gradient at the $\tau'=1$ surface.

Figure \ref{fig:schem} is an illustration of how the temperature gradient along the line of sight affects the observed polarization in the optically thick limit. The oblate grains are aligned with their short axes in the $x-$direction; therefore, the opacity in the $y-$direction is larger than that in the $x-$direction. By the Eddington-Barbier relation in the Rayleigh-Jeans limit, the observed intensity corresponds roughly to the temperature at the $\tau'=1$ surface. Since $\kappa_{x}$ is less than $\kappa_{y}$, the $\tau'_{x}=1$ location is further into the slab than the $\tau'_{y}=1$ location and thus the observed intensity in the $x-$ and $y-$directions are different based on three different cases of temperature gradient. Case B is when there is no temperature gradient. The observed intensity in the $x-$ and $y-$directions are the same, thus no polarization is produced. For Case A, temperature increases inwards (away from the observer) and has $\psi > 0$. Intensity in the $x-$direction comes from the hotter region while the intensity in the $y-$direction comes from the colder region closer to the surface. This results in polarization in the $x-$direction. Note that this is perpendicular to the direction for the optically thin case which is commonly used to interpret dust polarization data. On the other hand, temperature decreases inwards for Case C, or $\psi < 0$. The radiation polarized in the $y-$direction is stronger than that in the $x-$direction, thus resulting in polarization in the $y-$direction.

\begin{figure}
    \begin{center}
        \includegraphics[width=0.5\textwidth]{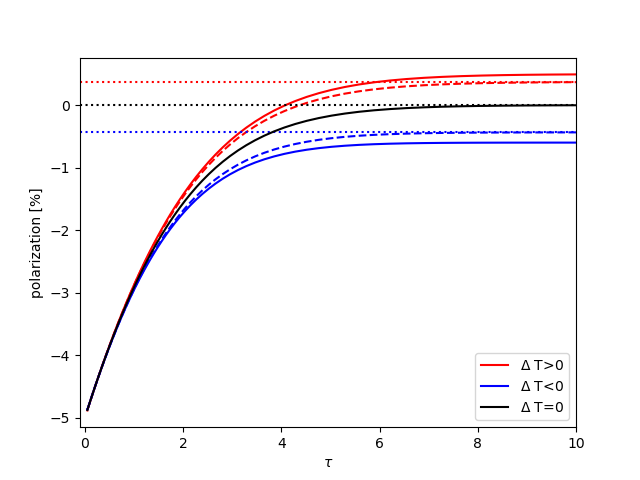}
    \end{center}
    \caption{
        Comparison of polarization fractions as a function of optical depth, $\tau$. The solid lines are the numerically obtained polarization fractions using the Planck function as the source function for various temperature gradients: $\Delta T = 0$ (isothermal, black), $\Delta T = -2$ (blue), and $\Delta T=2$ (red) in units of Kelvin per optical depth. $T_{0}$ is 25 K. The wavelength is 0.87 mm. The intrinsic polarization, $p_{0}$, is assumed to be $-5\%$. The dashed lines are the Rayleigh-Jeans counterpart of the solid lines of the same color, while the corresponding dotted horizontal lines are the asymptotic values in the optically thick limit given by Equation (\ref{eq:polpsi}). 
        }
    \label{fig:dTpol1D}
\end{figure}

We present a numerical example comparing polarization with and without temperature gradient in Figure \ref{fig:dTpol1D} for oblate grains. We have chosen a slab with temperature changing linearly with optical depth for illustration and $T_{0}=25$K, $\Delta T= \pm 2$K per unit optical depth, and $p_{0}=-5\%$. We integrate Equations (\ref{eq:Ix}) and (\ref{eq:Iy}) numerically with the Planck function as the source function $S$, focusing on a wavelength of 0.87 mm (ALMA Band 7, see Section \ref{sec:ObsModeling} below). We calculate the polarization for the slab with various total optical depths, $\tau$. This example illustrates clearly that although the temperature gradient ($\Delta T$) has relatively little effect for an optically thin slab (with $\tau \ll 1$), it controls both the magnitude and sign of the polarization for optically thick slabs. In particular, unlike the isothermal case, the polarization fraction does not go to zero in the non-isothermal cases as the total optical depth of the slab increases to infinity; it asymptotes to a finite value that is approximately the logarithmic temperature gradient at the $\tau'=1$ surface times $-p_{0}$ as given by Equation (\ref{eq:polpsi}). Note that the asymptotic value of the polarization fraction ($\vert p \vert$) is somewhat larger in the $\Delta T < 0$ case than the $\Delta T > 0$ case. This is because for a given surface temperature $T_{0}$, the temperature at the $\tau'=1$ surface is lower for the former than the latter making the asymptotic polarization fraction higher because it is proportional to the ratio of the temperature gradient $\Delta T$ (assumed to be fixed) to the temperature at $\tau'=1$ surface (see Equation (\ref{eq:poldTthick})).

Furthermore, we can also compare the polarization that is in the Rayleigh-Jeans limit. In this case, the slab with a surface temperature of 25 K has $h \nu / k T \sim 0.66$, which deviates significantly from the Rayleigh-Jeans limit. In both cases of temperature gradients, the magnitude of polarization is larger than its Rayleigh-Jeans counterpart. The trend can be understood as follows: polarization comes from the difference in the intensities from the colder and warmer $\tau'=1$ surfaces. The Planck function of the colder surface is further below the value given by the Rayleigh-Jeans approximation, while the deviation is less for the warmer surface, which increases the intensity from the warmer surface relative to the colder surface and thus the magnitude of the polarization fraction. 

From Equation (\ref{eq:poldTthick}), we recognize one interesting limit. In the case of a positive temperature gradient, $p/p_{0} \rightarrow -1$ when the gradient $\Delta T\rightarrow \infty$ or the surface temperature $T_0\rightarrow 0$. In other words, when temperature gradient is infinitely high or the surface temperature is 0, polarization is positive, i.e., in the $x-$direction, and has the same magnitude as the intrinsic polarization. Therefore, should one observationally identify polarization that is reversed from the intrinsic polarization, due to temperature gradient, the magnitude of polarization serves as a lower limit to the magnitude of the intrinsic polarization of the grain. 


\section{Effects of Temperature Gradient on Disk Polarization From Aligned Grains} \label{sec:ObsModeling}
Details of the temperature structure for protoplanetary disks remain poorly constrained observationally. However, one can calculate the expected temperature structure by radiative transfer given a dust opacity model, density distribution, and heating. In this section, we provide prescriptions of a fiducial disk model and simulate polarized emission from aligned grains to look for effects of temperature gradient. The emission is computed at $\lambda=0.87$mm, motivated by the fact that it is the highest frequency band (Band 7) at which ALMA can currently map dust polarization. Compared to other (lower) frequency bands, the dust emission at Band 7 has a higher optical depth, which makes it easier to show temperature gradient effects. 

\subsection{Dust Model} 
We consider a grain model similar to that of \cite{Pollack1994} which is comprised of $47\%$ water-ice, $30\%$ organics, and $23\%$ silicates by mass. The refractive indices are taken from \cite{Henning1996}\footnote{\url{https://www2.mpia-hd.mpg.de/home/henning/Dust_opacities/Opacities/opacities.html}} and opacity is calculated by the RADMC-3D Python implementation of Mie Theory \citep{Bohren1983}. For illustration purposes, we consider only grains of a relatively small size of $a=10 \mu m$, so that the effects of dust scattering, which are beyond the scope of this paper, can be ignored at the observing wavelength of 0.87mm. Our adopted dust model yields a rather small absorption opacity of $\kappa_{0.87\text{mm}} = 0.29$~cm$^{2}$ per gram of dust. For comparison, the oft-adopted opacity scaling from \cite{Beckwith1990} yields an opacity of $3.5$~cm$^2$ per gram of dust at 0.87mm for an opacity index $\beta=1$, which is similar to the value of $\sim 3.9$~cm$^2$ per gram of dust obtained through the DIANA project using also $a=10 \mu$m grains\footnote{\url{https://dianaproject.wp.st-andrews.ac.uk/data-results-downloads/fortran-package/}} \citep{Woitke2016}. In view of the potentially large uncertainty in the dust opacity, we prefer to characterize our model disk with the optical depth rather than the surface density (see the last paragraph of Section \ref{ssec:diskmodel} below), since the former is more directly relevant to the determination of the temperature structure, particularly for an actively accreting disk.

We have seen from Section \ref{sec:analytical} that one recovers the intrinsic polarization in the optically thin limit for oblate grains viewed edge-on. Generally though, polarization from thermal emission of non-spherical grains depends on the inclination of the axis of symmetry of the oblate grains with respect to the line of sight. In the electrostatic dipole approximation, suitable for small grains, the polarization fraction for oblate grains in the coordinate system described in Section \ref{sec:analytical} becomes
\begin{equation} \label{eq:ptheta}
    p(\theta) = \dfrac{p_{0} \sin^{2}(\theta)}{1 - p_{0}\cos^{2}(\theta)}
\end{equation}
where $\theta$ is the angle between the grain symmetry axis and the line of sight in the $xz-$plane \citep{LeeDraine1985, Yang2019}. Note that when $\theta=\pi/2$, oblate grains are viewed edge-on and one recovers the intrinsic polarization, $p=p_{0}$. When $\theta=0$, the grains are viewed face-on and there is no polarization even though grains have non-zero intrinsic polarization.

As an input to RADMC-3D, we prescribe the (absorption) opacities using Equation (\ref{eq:ptheta}). Specifically,
\begin{equation}
    p(\theta) = \dfrac{\kappa_{\nu, x}(\theta) - \kappa_{\nu,y}(\theta)}{\kappa_{\nu,x}(\theta) + \kappa_{\nu,y}(\theta)}
\end{equation}
where $\kappa_{x}$ and $\kappa_{y}$ are the opacities parallel and perpendicular to the grain alignment axis respectively. Since $\kappa_{\nu,y}$ does not vary with $\theta$, we further assume that $\kappa_{\nu,y} = \kappa_{\nu}$ where $\kappa_{\nu}$ is the opacity calculated from Mie Theory. With $p(\theta)$ from Equation (\ref{eq:ptheta}) and $\kappa_{\nu, y}$ known, $\kappa_{\nu,x}$ is also solved.

\subsection{Disk Model} \label{ssec:diskmodel}

Since the temperature gradient effect on the dust polarization along a given sight line depends on the sight line being optically thick (see Figure \ref{fig:schem}), it is expected to be more significant for more edge-on disks, the inner disk regions, and more massive disks. The last factor tends to favor younger disks that may be more embedded, as we will discuss in Section \ref{sec:Discussion} below). For such disks, one should in principle include an envelope, which could affect the disk temperature through ``back-warming" \citep[e.g.][]{Agurto2019}. However, for the purposes of testing and extending the 1D analytical model presented in the previous section, we will ignore such complications and adopt a generic viscous disk model with a disk surface density given by the self-similarity solutions of \cite{LyndenBell1974}:
\begin{equation} \label{eq:SurfaceDensity}
    \Sigma_{g}(R) = \Sigma_{c} \bigg( \dfrac{R}{R_{c}}\bigg)^{-\gamma} \exp{\bigg[ -\bigg( \dfrac{R}{R_{c}}\bigg)^{2-\gamma}\bigg]}
\end{equation}
where $R$ is the cylindrical radius and $R_{c}$ is the characteristic radius where exponential tapering becomes effective. The $\gamma$ parameter is the gas surface density exponent. $\Sigma_{c}$ is the characteristic surface density, which can be solved when given a disk mass, 
\begin{equation}
    \Sigma_{c} = (2-\gamma) \dfrac{M_{\text{disk}}}{2\pi R_{c}^{2}} \text{ .}
\end{equation}
We adopt a Gaussian distribution for vertical mass distribution:
\begin{equation}
    \rho_{g}(R,z) = 
        \dfrac{\Sigma_{g}}{\sqrt{2\pi}H} 
        \exp{ 
            \bigg[ 
                - \dfrac{1}{2} \bigg( \dfrac{z}{H}\bigg)^{2}
            \bigg]
            } 
\end{equation}
where $z$ is the vertical distance from the midplane and $H$ is the scale height, which is parameterized as $H=H_{t}(R/R_{t})^{q}$. The dust is assumed to be well-coupled to the gas with a gas-to-dust mass ratio of 100.  

Protoplanetary disks are illuminated by the central star. The irradiation impinges and heats the disk upper atmosphere causing a hotter atmosphere and a cold midplane. However, in accreting disks, viscosity leads to a production of heat preferentially near the midplane. Since we are interested in the temperature gradient, we consider two types of temperature structures: one with only stellar irradiation, termed the ``passive disk," and the other with accretion heating in addition to the stellar irradiation, termed ``accreting disk." For definitiveness, we will adopt a stellar mass of $M_{s}=1M_{\odot}$, a radius of $R_{s} = 2.5R_{\odot}$, and an effective temperature of 3900~K. These stellar parameters are similar to those inferred for IM Lup \citep{Cleeves2016}, but they are for illustrative purposes only. Other choices of stellar parameters should not qualitatively change the temperature structure of the irradiated disk, particularly the vertical increase of the temperature away from the disk midplane and the radial decrease of the temperature away from the central star. 

For the second case, accretion heating per unit volume in an accreting disk amounts to 
\begin{equation} \label{eq:qaccrho}
    q_{\text{acc}} = \rho_{g} \nu (R\dfrac{d\Omega}{dR})^{2}
\end{equation}
with $\nu$ as the viscosity coefficient and $\Omega=\sqrt{GM_{s}/R^{3}}$ is the Keplerian angular frequency (e.g. \citealt{DAlessio1998}). We use the standard $\alpha$-viscosity prescription and assume a steady-state accretion disk to relate viscosity to accretion rate:
\begin{equation}
    \dot{M} \equiv -2 \pi R \Sigma_{g} v_{r} = 3 \pi \Sigma_{g} \nu
\end{equation}
where the radial inward velocity $v_{r}=-3\nu / 2R$ is applied \citep{Shakura1973}. With gas density and accretion rate established, Equation (\ref{eq:qaccrho}) becomes
\begin{equation}
    q_{\text{acc}} = \dfrac{3\dot{M}}{4\pi} \dfrac{\Omega^{2}}{\sqrt{2\pi}H} \exp\bigg[ -\dfrac{1}{2} \bigg( \dfrac{z}{H} \bigg)^{2}\bigg] \text{ .}
\end{equation}
Accretion rate is set to be $5\times 10^{-8} M_{\odot}/\text{yr}$ which is broadly suitable for Class 0/I to Class II disks \citep[e.g.][]{Yen2017, Hartmann2006, Natta2006, Andrews2010, Mulders2017}. A higher accretion rate that may be expected for younger Class 0 disks would yield a higher midplane temperature, potentially making the temperature gradient effect on the dust polarization even more pronounced.

With dust density, dust opacity, and heating mechanisms provided, temperature is calculated using RADMC-3D. We ignore grain alignment when calculating temperature, since polarization fraction of the grain is small and thus will not severely affect the radiation field. As a fiducial disk model, we set $\gamma=0.5$, $R_{c}=50$ AU, $H_{t}=5$AU, and $q=1.125$. The synthetic observations of these disks are done in four different disk inclinations: $0^{\circ}$(face-on), $45^{\circ}$, $75^{\circ}$ and $90^{\circ}$(edge-on). The optical depth of a given disk depends on the disk inclination. A disk that is optically thick when seen edge-on may be entirely optically thin face-on and temperature gradient effects may not be seen. For the purpose of illustrating the temperature gradient effects even in the face-on case, instead of providing a fixed value for disk mass, we opted to vary the disk mass at each inclination such that the optical depth is 5 at a distance of 50AU from the central star along the major axis
\footnote{Assuming the standard gas-to-mass ratio of 100, the edge-on disk has a total gas mass of $0.085 (3 [{\rm cm}^{2} / {\rm g}] / \kappa_{\rm 0.87mm})$~M$_\odot$, corresponding to a minimum Toomre $Q$ value of about $2.1 (3 [{\rm cm}^{2} / {\rm g}]/ \kappa_{\rm 0.87mm})^{-1}$ in the accreting case; it would be gravitationally stable if the opacity at 0.87mm is close to or larger than $3 {\rm cm}^2$ per gram of dust \citep{Beckwith1990, Woitke2016}. As expected, the face-on disk needs a much higher mass of $0.55 ( 3 [{\rm cm}^{2} / {\rm g}]/ \kappa_{\rm 0.87mm})$~M$_\odot$ to reach the same optical depth of 5 at 50 AU. It is likely gravitationally unstable, since its minimum Toomre $Q$ value in the accreting case is about $0.33 (3 [{\rm cm}^{2} / {\rm g}]/ \kappa_{\rm 0.87mm})^{-1}$. An implication is that if polarization reversal is observed in a face-on disk on the 50 AU scale modeled here, the disk must be exceptionally massive. This may indeed be the case for the low-mass Class 0 protostar NGC 1333 IRAS 4A1 (see discussion in Section \ref{sec:MultiBand} below), for which \cite{SeguraCox2018} estimated an exceptionally large disk mass in the range of $0.85 - 1.9$~M$\odot$ (see their Table 3). The required disk mass would be reduced if polarization reversal is detected within a smaller radius around the central star.}; 
the value of optical depth is motivated by Section \ref{sec:analytical} (see Figure \ref{fig:dTpol1D}), where temperature gradient effects become dominant at $\tau \gtrsim 5$. For illustration, we assume oblate grains with an intrinsic polarization of $p_{0}=-5\%$ and the grains are aligned with their short axes parallel to a toroidal magnetic field in the disk. 

\subsection{Results for the Passive Disk} \label{ssec:passive_results}

\begin{figure*}
    \centering
    \includegraphics[width=0.8\textwidth]{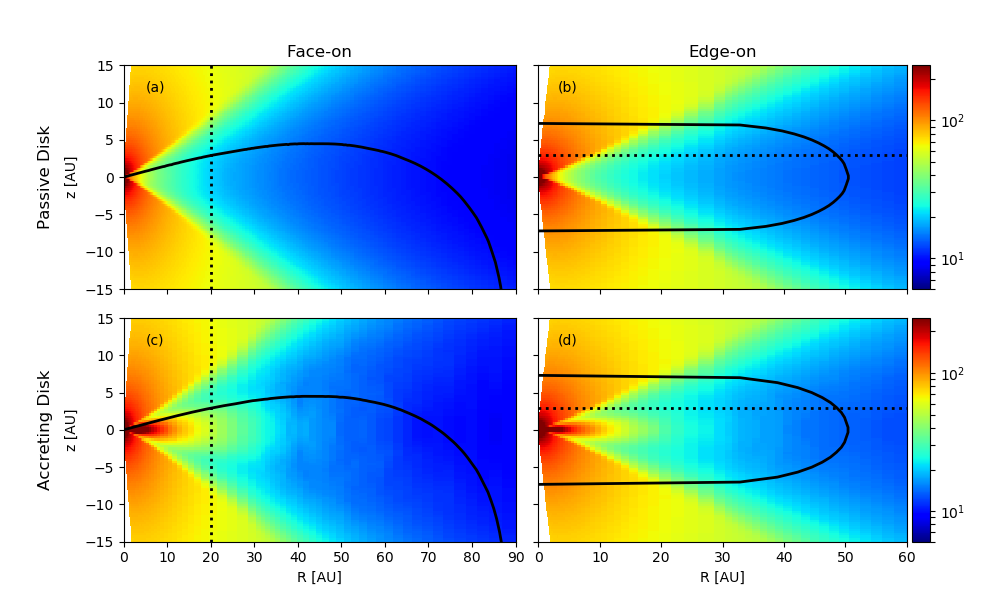}
    \caption{
            Colored plots depicting the calculated temperature structure as a function of radius and height in a cylindrical coordinate system (R,z) on a meridian plane of the axisymmetric disk. Disk midplane corresponds to $z=0$. The four different disk models are: (a) face-on passive disk, (b) edge-on passive disk, (c) face-on accreting disk, and (d) edge-on accreting disk. The black solid lines in the left panels mark the surface where the $\tau'=1$ as viewed from above the disk (i.e., face-on), while those in the right panels correspond to the $\tau'=1$ surface as viewed along the disk plane (from right to left, i.e., edge-on). The black dotted lines mark two of the example sight lines discussed in Sections \ref{ssec:passive_results} and \ref{ssec:accretion_results}. Note that the horizontal range is different since we are interested in seeing the location of the $\tau'=1$ surface. 
        }
    \label{fig:T_structure}
\end{figure*}

This subsection presents the disk with only stellar irradiation as the source of heating. Its temperature structure is shown in the upper panels of Figure \ref{fig:T_structure}. Stellar irradiation impinges the surface of the disk. Grains in the atmosphere see the incoming radiation and re-radiate only a portion of the heat into the disk midplane. Grains in the midplane of the disk where it is optically thick do not receive as much heating and become coldest in the midplane. This creates the typical structure of a hotter atmosphere and a cold midplane which was already seen with ALMA \citep[for example, the ``hamburger" morphology of HH 212 in][]{LeeCF2017_lane}. Note that the disk masses are modified for different inclinations as mentioned in Section \ref{ssec:diskmodel} and thus the calculated temperature structure is also different.

\begin{figure*} 
    \centering
    \includegraphics[width=\textwidth]{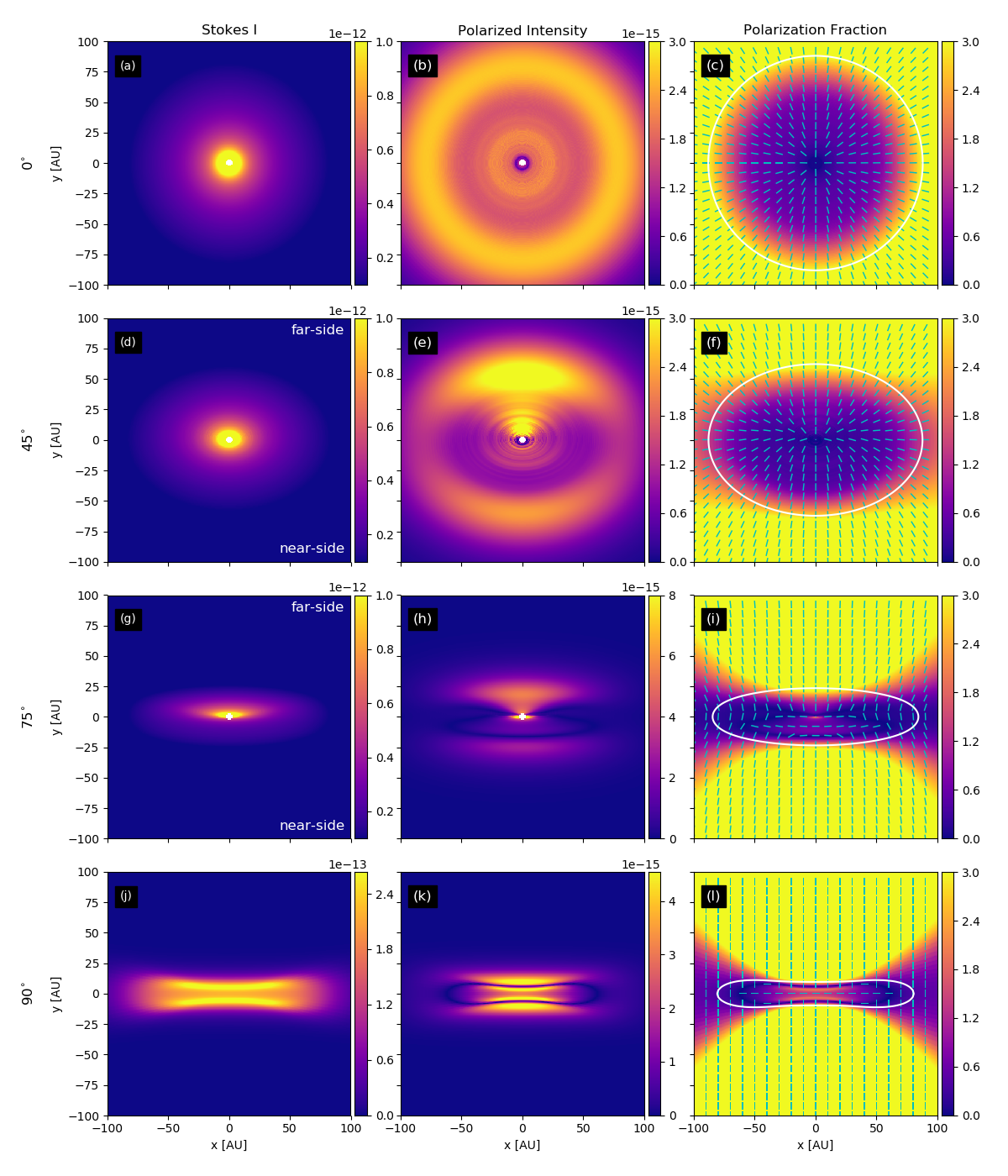}
    \caption{
        The modeled images of the disk with only stellar irradiation at various disk inclinations. The rows correspond to inclinations of $0^{\circ}$, $45^{\circ}$, $75^{\circ}$, and $90^{\circ}$, and the columns correspond to Stokes I, polarized intensity, and absolute polarization fraction with overplotted E vectors have the same length (i.e., not proportional to the polarization fraction). Stokes I and polarized intensity are in units of $\text{erg}\cdot \text{s}^{-1} \cdot \text{sr}^{-1} \cdot \text{cm}^{-1} \cdot \text{Hz}^{-1}$ while polarization fraction is in percentage. The color scale for the Stokes I and polarized intensity vary for each case to highlight the main features. The polarization fractions are plotted to the same scale. The white contour in the polarization fraction plots (right column) encloses the region where the total optical depth exceeds unity. 
    }
    \label{fig:Tirr_obs}
\end{figure*}

\begin{figure*}
    \centering
    \includegraphics[width=\textwidth]{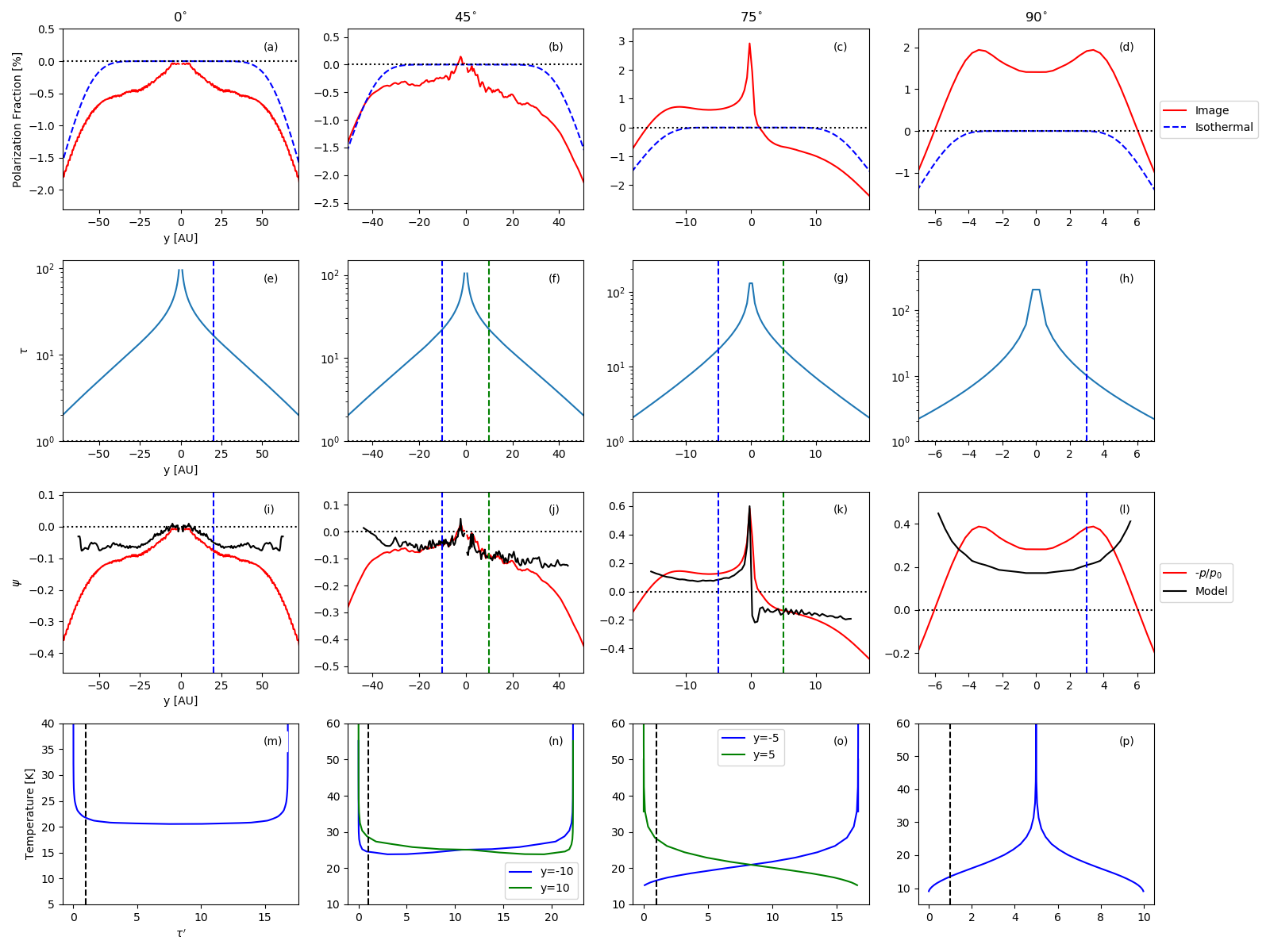}
    \caption{
        The cuts along minor axis (x=0) for the four different disk inclinations by column. The columns starting from the left correspond to $0^{\circ}$, $45^{\circ}$, $75^{\circ}$, and $90^{\circ}$ inclinations. The first row compares the polarization fraction from the Figure \ref{fig:Tirr_obs} (red solid lines) and the polarization fraction for the isothermal case numerically computed using Equation (\ref{eq:poliso}) (blue dashed-lines). The second row shows the optical depth. The third row compares the temperature gradients. The logarithmic temperature gradient, $\psi$, directly extracted from the model is in black and only plotted where the optical depth is greater than three. The logarithmic temperature gradient inferred from Equation (\ref{eq:polpsi}) based on $-p/p_{0}$ is plotted in red. Note that the ranges for both the horizontal and vertical axes are different for the three cases. The bottom panels are example temperature profiles as a function of optical depth at various lines of sight. $\tau'=0$ is the surface closest to the observer. The color of the profile corresponds to the vertical dashed lines in the second and third row where it marks the location of the line of sight. For example, the blue temperature profile in panel o is extracted at $-5$ AU along the minor axis, or the blue dashed line in panel k, and accordingly for the green line. The black dashed vertical lines for the bottom panels denotes $\tau'=1$. 
    }
    \label{fig:Tirr_minor}
\end{figure*}

The images of Stokes I, polarized intensity, and polarization fraction with polarization direction are shown in Figure \ref{fig:Tirr_obs}. Polarization vectors point along the projected major axis of the oblate grains in the isothermal and optically thin limits. With the minor axis aligned toroidally for all grains, one can quickly verify that the vectors should point radially for the face-on case and point along the minor axis of the projected disk (in the $y-$direction of the image) for the edge-on case. This is viewed in the outermost regions of the image. These polarization vectors outside of the region where optical depth becomes unity (see panels c, f, i, and l), labeled by the white contour, are not affected by the temperature structure. Within the region where optical depth is greater than unity, the temperature gradient starts to become an important factor for the polarized intensity and direction. To be more quantitative, we compare the polarization along the $y-$direction (the minor axis for inclined disks) for each inclination to that for the isothermal case (Equation (\ref{eq:poliso})) in the top panels of Figure \ref{fig:Tirr_minor} and the optical depth from the model is plotted in the second row of Figure \ref{fig:Tirr_minor}. The near-side of the disks with $45^{\circ}$ and $75^{\circ}$ inclinations is where $y$ is negative. 

For the face-on case, in panel c of \ref{fig:Tirr_obs}, the polarization fraction decreases towards the center of the disk and the vectors remain in the radial direction. These are reminiscent of polarization by direct emission in the isothermal case. In the isothermal case, polarization quickly becomes zero in the optically thick regions ($\tau \gtrsim 6$) as shown in Figure \ref{fig:dTpol1D}. This is not what we find in the disk model, where the polarization fraction remains significant even in the optically thick region shown in panel a of Figure \ref{fig:Tirr_minor}. The remaining negative polarization fraction is due to the hotter atmosphere viewed in front of the cold midplane, which corresponds to Case C in Figure \ref{fig:schem}, and results in a finite polarization along the direction of the long axis of the grain projected in the plane of the sky (i.e., the direction of $p_{0}$) even in a very optically thick region).

\begin{figure}
    \centering
    \includegraphics[width=\columnwidth]{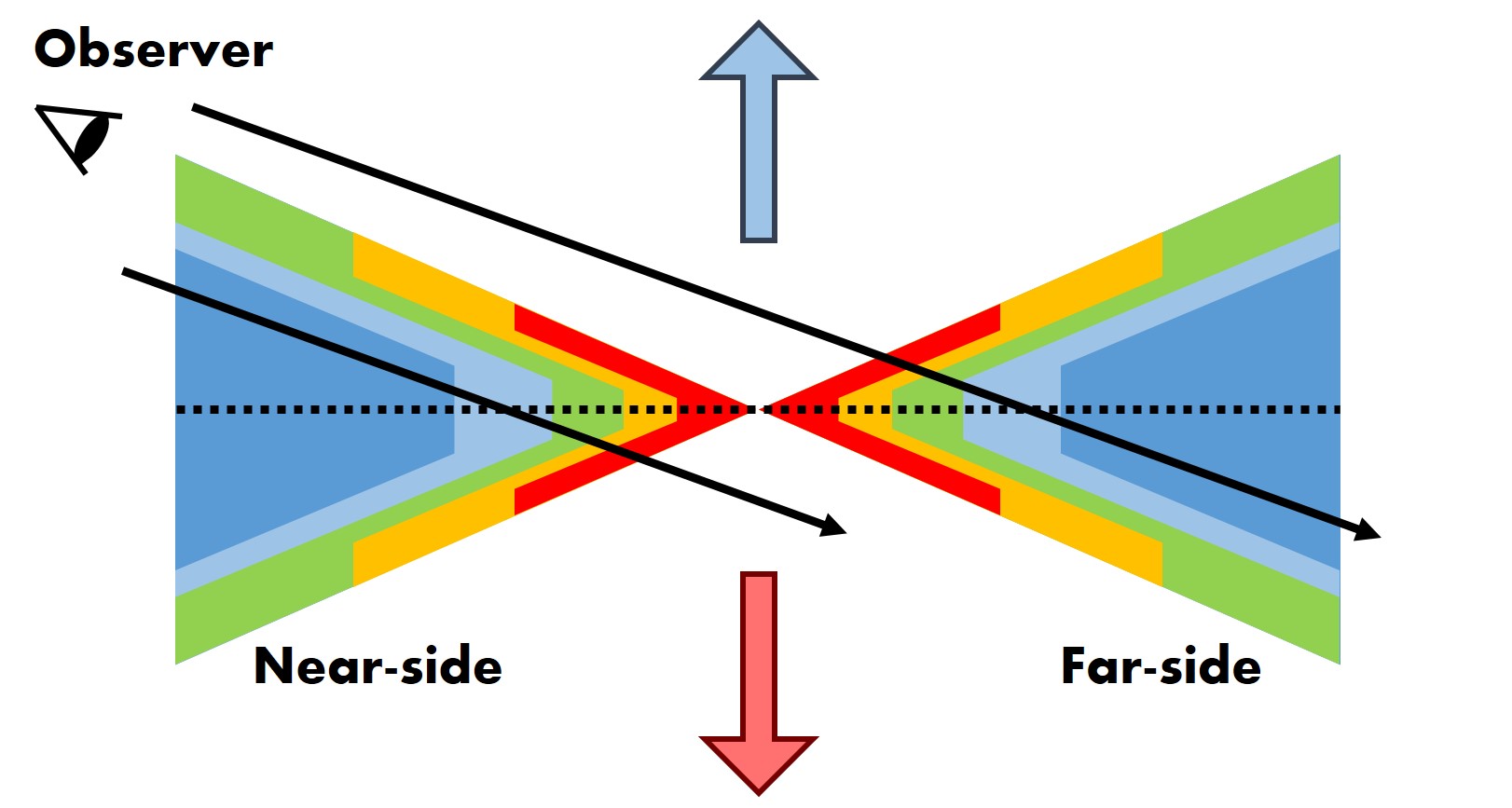}
    \caption{
            Schematic drawing of different temperature gradients traced by different line of sights along the minor axis for the $75^{\circ}$ inclination case. The dumbbell shaped region represents the meridional temperature structure of the disk, where the color scheme is used in the same sense as Figure \ref{fig:T_structure} with temperature decreasing from red to blue. The dotted line is the disk midplane. The observer is located in the top left corner. The black arrows represent different line of sights, one crosses the near-side and the other crosses the far-side. The red and blue arrows perpendicular to the midplane represent the red-shifted and blue-shifted jet of the disk if it exists.
        }
    \label{fig:nearfar}
\end{figure}

Now we consider the other extreme, which is the edge-on case (the bottom row of Figure \ref{fig:Tirr_obs}). In the optically thin regions, polarization vectors point in the $y-$direction of the image, which is commonly expected for grains aligned with their short axes along a toroidal magnetic field. Unlike the face-on case, there is a strong azimuthal variation in polarization fraction even in the optically thin regions (outside the white contour in panel l). Grains located on the minor axis of the disk are viewed more edge-on than those located on the major axis (in the outer optically thin region), and hence produce a more polarized emission. More interestingly, in the inner optically thick region, the polarization vectors rotate at an angle of $90^{\circ}$, and become perpendicular to the optically thin regions. This $90^{\circ}$ rotation comes about because grains are aligned with their short axes parallel to the disk midplane, so that the radiation polarized along the midplane has a smaller opacity and thus comes from a smaller radius where the temperature is higher, resulting in a net polarization along the short (rather than the long) axis of the grain in the plane of the sky, analogous to Case A of Figure \ref{fig:schem}. Panel d of Figure \ref{fig:Tirr_minor} shows that the signed polarization fraction becomes positive (i.e. a polarization reversal relative to $p_{0}$) which is completely different from the face-on case where the signed polarization fraction remains negative (i.e. in the same direction as $p_{0}$).

The $75^{\circ}$ inclination case is the highly inclined but not quite edge-on case. The oblate grains remain more edge-on along the minor axis of the disk, but less so in the outer edge of the disk. Thus the pattern for polarization fraction in the optically thin regions is similar to that of the $90^{\circ}$ inclination (edge-on) case. Within regions where the total optical depth exceeds unity, the near-side (bottom half of the image) shows polarization reversal like the edge-on case and the far-side maintains the excess polarization like the face-on case (compared to the isothermal case). This near-far side asymmetry comes about because the dust is assumed to have a finite (angular) thickness and the disk surface on the far-side is viewed more face-on than the near-side\footnote{The same geometric effect is the reason behind the near-far side asymmetry in scattering-induced polarization in optically and geometrically thick disks \citep{Yang2017_nearfar}.}. The line of sight in the near-side travels towards decreasing radii which generally leads to positive temperature gradients whereas that in the far-side has a negative temperature gradient making the near-side resemble more closely the edge-on case and the far-side the face-on case. Figure \ref{fig:nearfar} is a schematic drawing to illustrate the differences of the temperature gradient traced by different line of sights. The line of sight that intercepts the near-side goes through a region with decreasing temperature and vice versa for the far-side. This difference in positive or negative temperature gradient causes polarization reversal in the near-side and none in the far-side. This is clearly seen in panel c of Figure \ref{fig:Tirr_minor} where the polarization in the near-side is positive and resembles the edge-on case and vice versa for that in the far-side. 


The $45^{\circ}$ inclination case serves as the transition from the face-on to the $75^{\circ}$ inclination case. The polarization fraction seen in Figure \ref{fig:Tirr_obs}f broadly resembles that of the face-on case in Figure \ref{fig:Tirr_obs}c, because its temperature profiles along the line of sight and the optical depths are similar (compared to the $75^{\circ}$ and edge-on cases). The asymmetry along the minor axis plotted in Figure \ref{fig:Tirr_minor}b is caused by the same mechanism as in the $75^{\circ}$ case, but less dramatic.

The top panels of Figure \ref{fig:Tirr_minor} show the important point that the isothermal case commonly adopted for interpreting dust polarization data on relatively large scales (e.g. molecular clouds) fails for the optically thick parts of the disk, where the temperature varies strongly both radially and vertically. Specifically, the polarization fraction does not go to zero as the optical depth increases. It asymptotes to a finite value depending on the temperature gradient along a given line of sight, as we stressed earlier using the analytical 1D slab model (Figure \ref{fig:dTpol1D}). To help interpret the numerical results, we make use of the analytical 1D slab model presented in Section \ref{sec:analytical} and compare in the third row of Figure \ref{fig:Tirr_minor} the two quantities, $-p/p_{0}$ and $\psi$, that appear in Equation (\ref{eq:polpsi}) along the minor axis. We obtain $\psi$ at the $\tau'=1$ surface along each line of sight. The latter quantity is plotted only for the optically thick lines of sight where the total optical depth $\tau$ is greater than three. Along such optically thick sight lines, the negative of the normalized polarization fraction $p/p_{0}$ does indeed follow $\psi$ qualitatively, which lends support to the basic physical picture outlined in Section \ref{sec:analytical}. 

Quantitatively, there is some difference between $-p/p_{0}$ and $\psi$, however. This difference is to be expected, since the two are exactly equal only when the observing wavelength $\lambda$ is in the Rayleigh-Jeans tail of the black body radiation everywhere along the line of sight, the temperature gradient with respect to the optical depth, $dT / d\tau'$, is constant, and the total optical depth is infinite. In the bottom panels of Figure \ref{fig:Tirr_minor}, we plot the temperature as a function of optical depth along the line of sight, $\tau'$, at one representative sight line for the face-on and edge-on cases (these sight lines are shown as dotted lines in Figure \ref{fig:T_structure}) and at two sight lines for the $45^{\circ}$ and $75^{\circ}$ inclination cases. The dashed lines mark where optical depth becomes 1. We can see that the temperature near the $\tau'=1$ surface is relatively low (of order 20 K), which leads to a significant deviation at ALMA Band 7 from the Rayleigh-Jean approximation adopted for deriving Equation (\ref{eq:polpsi}). As explained in Section \ref{sec:analytical} and shown in Figure \ref{fig:dTpol1D}, the deviation leads to a larger polarization fraction based on the full Planck function ($\vert p/p_{0} \vert$) compared to that estimated under the Rayleigh-Jeans approximation ($\vert \psi \vert$), which explains, at least in part, the difference between the two quantities plotted in the third row of Figure \ref{fig:Tirr_minor}. The second condition does not strictly hold either in the stellar irradiated disk under consideration, since temperature gradient is obviously not constant for these sight lines. With the sight lines for the face-on case as an example, temperature rapidly decreases from $\tau'=0$ to $\tau'=1$. This is effectively a larger temperature gradient than that directly extracted at $\tau'=1$ and thus increases the difference between the intensities along the $x-$ and $y-$direction which produces a larger polarization, consistent with the results shown in the third row of Figure \ref{fig:Tirr_minor}. Despite these, the difference between the normalized negative polarization fraction $-p/p_{0}$ and the logarithmic temperature gradient $\psi$ is relatively moderate, typically by a factor of 2 or less, indicating that polarization from aligned grains can indeed be used to determine the temperature gradient to a reasonable accuracy in the optically thick part of a disk. 

\subsection{Results for the Accreting Disk} \label{ssec:accretion_results}

\begin{figure*}
    \centering
    \includegraphics[width=\textwidth]{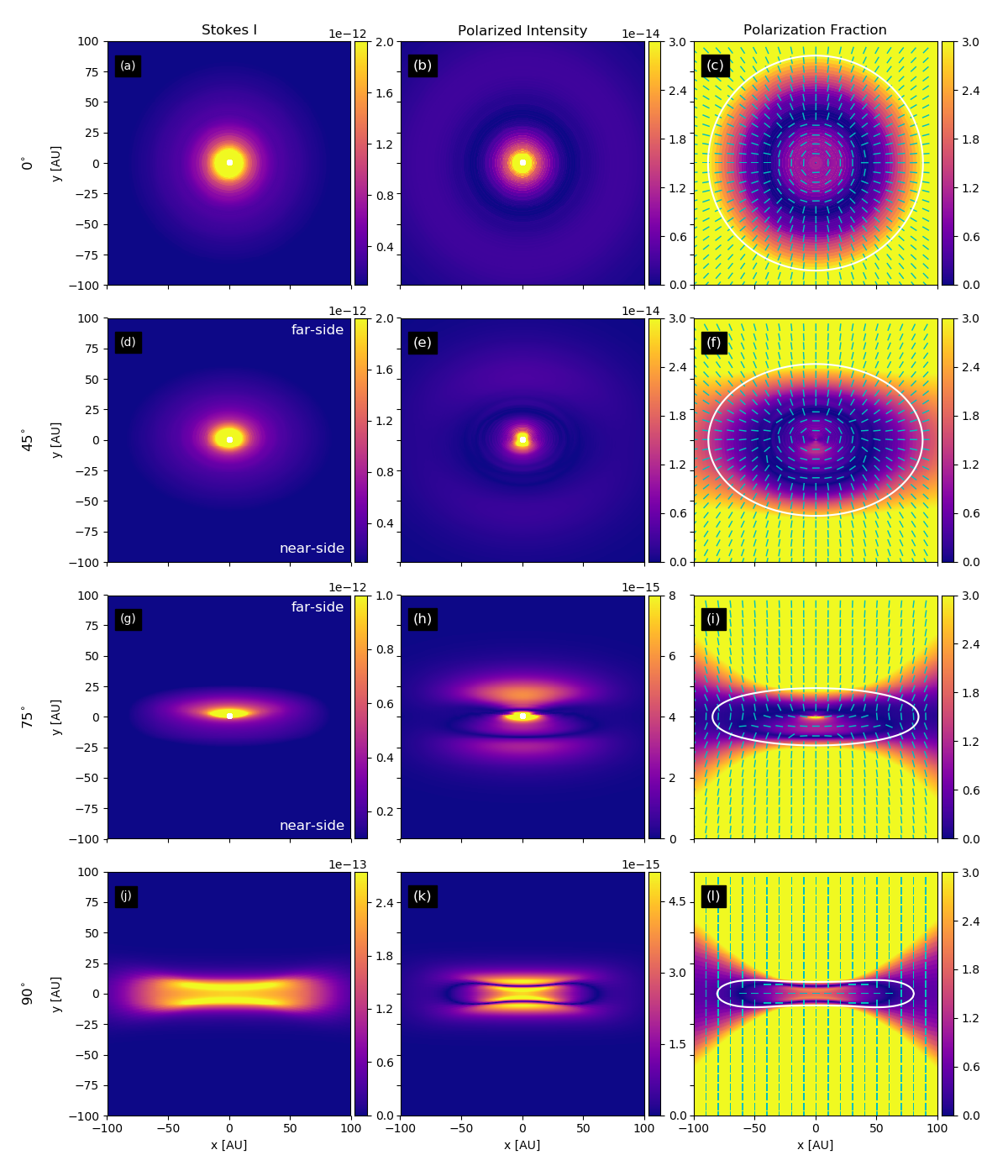}
    \caption{
    The modeled images for the disk including accretion heating at various inclinations. The figure is presented in the similar manner as Figure \ref{fig:Tirr_obs}. 
    }
    \label{fig:Tacc_obs}
\end{figure*}

\begin{figure*}
    \centering
    \includegraphics[width=\textwidth]{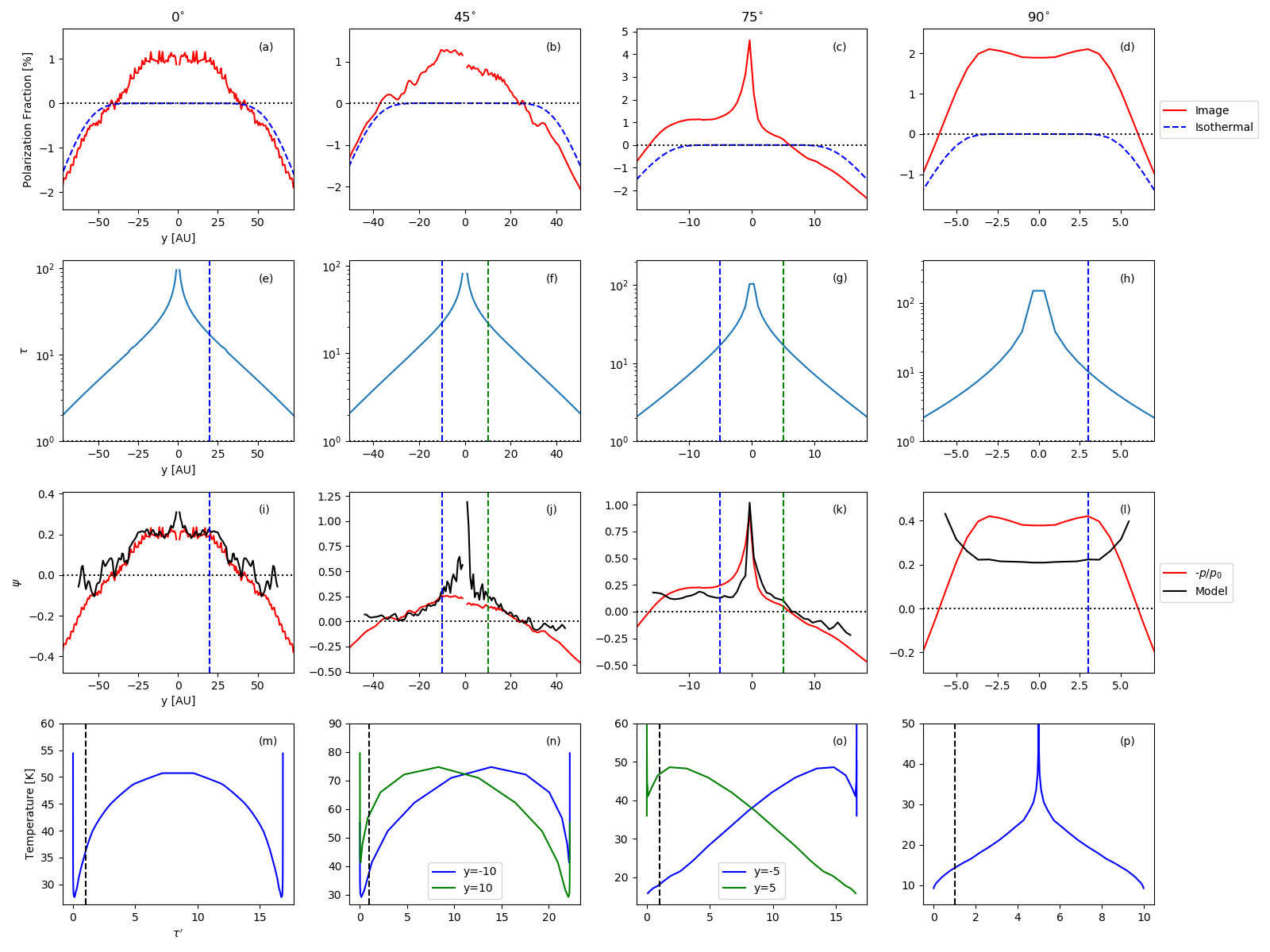}
    \caption{
    The cuts along the minor axis for the disk including accretion heating. The figure is presented in the same way as Figure \ref{fig:Tirr_minor}. 
    }
    \label{fig:Tacc_minor}
\end{figure*}

Viscous heating is proportional to gas density; therefore, in the regions where gas density is larger (near the midplane in this case), grains can be heated more. The temperature structure is shown in the bottom panels of Figure \ref{fig:T_structure}. The atmosphere is still directly heated by the star. The midplane on the other hand becomes hotter and the temperature no longer monotonically decreases to the midplane\footnote{We note that the viscous heating has a strong effect on the midplane temperature out to a larger radius in our model than in the model shown in Figure 2 of \cite{Bitsch2015}, even though the heating rate is similar in the two cases. The difference likely comes from a higher Rosseland mean optical depth vertically through the disk in our model than in theirs at a given radius, which leads to a higher midplane temperature, though the optical depth from \cite{Bitsch2015} is not readily known.}. The resulting images are shown in Figure \ref{fig:Tacc_obs} and the cuts along the minor axis are plotted in Figure \ref{fig:Tacc_minor}. Both figures are plotted in the same way as Figure \ref{fig:Tirr_obs} and Figure \ref{fig:Tirr_minor} respectively.


Starting from the face-on case, the polarization vectors in panel c of Figure \ref{fig:Tacc_obs} show a striking difference from that of sole stellar irradiation in panel c of Figure \ref{fig:Tirr_obs}. Polarization fraction can be reversed even in the face-on case because of the temperature structure. This can be verified in Figure \ref{fig:T_structure}c where the $\tau'=1$ surface (viewed vertically downward from above) is plotted on the meridionial plane for the face-on case. In regions within 40AU, the temperature gradient responsible for polarization is positive and thus produces polarization reversal (Figure \ref{fig:Tacc_minor}a). Similar to the face-on case, the $45^{\circ}$ inclination case also hosts an inner region where the polarization is reversed. 

Polarization fraction for the edge-on case is relatively similar to the disk with only stellar irradiation. Inclusion of accretion heating does not drastically change the temperature structure near the $\tau'=1$ surface, which is located at relatively large radii, where the gravitational binding per unit mass of material is low and the accretion heating is relatively inefficient. One can verify the similarities of the temperature structure traced by the $\tau'=1$ surface in panels b and d of Figure \ref{fig:T_structure}. From Figure \ref{fig:Tacc_minor}c, we can see that within the optically thick regions, the highly inclined case of $75^{\circ}$ inclination has polarization reversal in the near-side and exhibits the reversal in the far-side as well between $y=0$ and $5$ AU. Beyond $y\sim5$AU of the far-side, the temperature gradient along the line of sight becomes negative, since the line of sight begins to leave the warmer regions of the midplane and the polarization fraction becomes similar that of the passive disk (Figure \ref{fig:Tirr_minor}c). 

From the third row of Figure \ref{fig:Tacc_minor}, we can see that $-p/p_{0}$ traces $\psi$ for the accreting disk, just as the passive disk. The main difference is in the face-on case, where the deviation between $-p/p_{0}$ and $\psi$ are smaller than that of the passive disk. For the most part, this is because the temperature is higher in the accreting disk case than its passive counterpart (compare, e.g., Figure \ref{fig:Tirr_minor}g to Figure \ref{fig:Tacc_minor}g), so the Rayleigh-Jeans approximation is more accurate. Furthermore, as a typical sight line passes through the atmosphere and reaches the midplane, the temperature rapidly decreases and then increases again as shown in Figure \ref{fig:Tacc_minor}g. The effective temperature gradient near the disk surface averages out to a value that is similar to that at the $\tau'=1$ surface. The non-Rayleigh-Jeans effect can be seen clearly in the $75^{\circ}$ case, where the temperature of the far-side (positive $y$ location along the minor axis) is greater than that of the near-side at the $\tau'=1$ region (see Figure \ref{fig:Tacc_minor}k). Thus, the deviation between $-p/p_{0}$ and $\psi$ is smaller for the far-side.

\section{Discussion} \label{sec:Discussion}

Our main result is contained in Equation (\ref{eq:polpsi}): temperature gradient affects polarization when optically thick and the normalized polarization fraction $-p/p_0$ is related to its logarithmic temperature gradient $\psi\equiv d\ln T/d\tau' \vert_{\tau'=1}$ along the line of sight. In principle, polarization observations allows us to infer both the magnitude and direction of the temperature gradient, which is an intrinsic property of the disk, or any other density distribution (e.g. envelope). In practice, there are several complications for applying this technique to infer the temperature gradient. First, the observed polarization must be emitted by aligned grains, which is difficult to ascertain in general because it can also be produced by scattering \citep{Kataoka2015, Yang2016_HLTau}.
Second, the relation applies only to optically thick regions, but optical depth is hard to determine because of large uncertainties in dust opacity. A third difficulty is that the intrinsic polarization fraction, $p_0$, must be determined first, which may not be easy to do. These challenges need to be overcome in order to apply Equation (\ref{eq:polpsi}) to infer the disk temperature gradient from polarization observables.

\subsection{The Role of Multi-Wavelength Polarization Observations} \label{sec:MultiBand}

To illustrate some of these practical complications, let us consider the deeply embedded Class 0 protostar NGC1333 IRAS4A1. It is one of the brightest protostellar continuum sources on the (100 au) inner envelope to disk scale at both (sub)millimeter and centimeter wavelengths \citep{Looney2000, Cox2015}. It should therefore have one of the highest optical depths, making it an ideal target for applying our technique. Just as importantly, spatially resolved dust polarization has been detected at both centimeter \citep{Cox2015, Liu2016} and (sub)millimeter wavelengths \citep{Ko2020}. The orientations of the (sub)millimeter ($\sim 0.87$ and $1.3$ mm) polarization are rotated by nearly $90^\circ$ from those at centimeter wavelengths in the brightest disk-like structure. This rotation is consistent with the polarization reversal expected between the polarization emitted by aligned grains at a wavelength that is optically thin or moderately optically thick (centimeter) and that at a wavelength that is very optically thick (sub/millimeter). If this interpretation is correct, we can immediately conclude that there must be a temperature increase into the disk-like structure around the (sub)millimeter $\tau'=1$ surface (see Case A in Figure \ref{fig:schem}). This conclusion is supported by the detection of H$_2$CO \citep{Su2019} and CH$_3$OH \citep{Sahu2019} absorption lines, which requires a colder (molecule-bearing) layer in front of a warmer background. If the disk-like structure is viewed relatively face-on, as suggested by the relatively small inclination angle of $\sim 35^\circ$ ($0^\circ$ means exactly face-on) inferred by \cite{SeguraCox2018} through detailed modeling of the dust morphology at 8~cm, the (positive) gradient would point to a temperature increase towards the midplane, with the implication that the disk-like structure is heated internally (rather than only externally by, e.g., stellar irradiation), possibly by active accretion, as illustrated in Figure \ref{fig:Tacc_obs}f. 

To estimate the (positive) temperature gradient in IRAS 4A1 more quantitatively, one would need to determine the intrinsic polarization fraction $p_0$ at the (sub)millimeter wavelengths (where the polarization is reversed) first. One possibility is to adopt the polarization fraction at the same wavelength in a region that is known to be optically thin. However, this is difficult to do for IRAS 4A1 because the optically thin regions appear to be located in the protostellar envelope, where the grain properties and alignment efficiency may be quite different from those on the disk scale. Another possibility is to adopt the polarization fraction in the same region but at a wavelength where the dust is optically thin (or at most moderately optically thick) under the assumption that the intrinsic polarization fractions are the same at different wavelengths (which may or may not be true). For IRAS 4A1, one such wavelength is 8~cm, where the average polarization fraction is observed to be of order $15\%$ \citep{Cox2015, Liu2016}. If we take this value as an estimate of the intrinsic polarization fraction at 0.87~mm (i.e., $p_0\approx -15\%$), we obtain, from Equation (\ref{eq:polpsi}) and the average polarization fraction of $p\sim 3\%$ in the region of reversed polarization \citep{Ko2020}, a rough estimate for the logarithmic temperature gradient $\psi\equiv d\ln T/d\tau'\vert{_{\tau'_{0.87 \text{mm}}=1} }\sim 0.2$. In other words, the temperature would double over an optical depth interval of $\Delta \tau' \sim 5$.


%
%

IRAS 16293B is another deeply embedded Class 0 protostar with a nearly face-on ``candidate" disk and has spatially resolved polarization data at both centimeter and (sub)millimter wavelengths \citep{Liu2018, Rao2014, Sadavoy2018}. Interestingly, the polarization orientations at 1.3~mm and 6.9~cm are consistent with each other and the polarization fractions in the two wavebands are comparable (\citealt{Sadavoy2018}, see their Figure 9); the situation is a stark contrast to the NGC1333 IRAS4A1 case discussed above. If the polarization is produced by aligned grains in both wavebands, the consistent orientations would imply (1) that there is no polarization reversal in either waveband or (2) that there is reversal in both. Given that the disk is almost surely optically thick at 1.3~mm \citep{Sadavoy2018}, the former scenario would imply that any temperature increase towards the disk midplane, if present\footnote{Red-shifted absorption is detected in CH$_3$OCHO \citep{Pineda2012} and CH$_3$OH \citep{Zapata2013}, indicating a colder molecule-bearing layer in front of a warmer background. However, the colder layer is moving at a substantial speed ($\sim 0.5-0.7$~km/s) towards the disk, making it unlikely part of the disk proper.}, is not fast enough to overwhelm the intrinsic polarization and produce reversal. In this case, the 1.3~mm emission is expected to have a much smaller polarization fraction $p$ than that at (the much more optically thin) 6.9~cm, unless its intrinsic polarization fraction ($p_0$) is much larger; such an increase in $p_0$ may offset the optical depth effects to produce a polarization fraction at 1.3~mm comparable to that at 6.9~cm, as observed. In the second scenario where the polarization is reversed in both wavebands, the polarization fraction $p$ would be set by the (positive) temperature gradient $\psi$ near the $\tau'=1$ surface and the intrinsic polarization fraction $p_0$ for the respective wavelength. The fact that $p$ is observed to be comparable means that the product $\psi p_0$ is similar in both wavebands according to Equation (\ref{eq:polpsi}). One way to distinguish between the two possibilities is to independently determine the optical depth at 6.9~cm, which is required to be substantially optically thick for the second scenario.
%
%

A third case of interest is the more evolved, Class I disk BHB 07-11. Its polarization is detected in three ALMA Bands (0.87, 1.3 and 3~mm), with orientations that are consistent with one another in all three bands and fractions that increase with wavelength \citep{Alves2018}. Since the 3~mm emission is likely optically thin in this relatively evolved disk, its polarization is unlikely reversed. The consistency in multi-band polarization orientations then implies that the polarization is not reversed at the two shorter wavelengths either, despite their higher optical depths. The implication is that the optical depth is not high enough and/or the temperature does not increase fast enough along the line of sight to reverse the intrinsic polarization even at the shortest wavelength (0.87~mm). Indeed the brightness temperature decreases with longer wavelength and also the presented spectral index map (see their Figure 5) has values greater than 2, which are indicative of low optical depth\footnote{Optically thick regions can also have spectral indices greater than 2 if they are dominated by scattering with an albedo increasing with wavelength \citep{Zhu2019}, although polarization for BHB 07-11 is likely not due to self-scattering \citep{Alves2018}} and/or decreasing temperature gradient. The decrease of polarization fraction with decreasing wavelength is qualitatively consistent with the expected larger optical depth effects at shorter wavelengths. Whether different optical depths can quantitatively explain the observed difference in polarization fraction remains to be determined, however.
%
%

\subsection{Nearly Edge-On Disks}  \label{sec:EdgeOn}

Nearly edge-on sources are particularly useful targets to investigate the effects of the temperature gradient on polarization from aligned grains, because the edge-on view maximizes the optical depth along the line of sight, which passes through disk regions of different radii, where the temperatures are expected to be different. So far, there are two Class 0, L1527 \citep{Harris2018} and HH 212 \citep{LeeCF2018}, and one Class I, HH 111 \citep{LeeCF2018}, disks that have nearly edge-on orientations and spatially resolved continuum polarization observations by ALMA, but thus far each source has only been observed at one wavelength with ALMA. Nevertheless, we will discuss them one by one below.  

%
%
%
%

With an inclination angle of $\sim 85^\circ$, L1527 in the Taurus molecular cloud is one of the closest Class 0 protostars with a nearly edge-on disk \citep{Tobin2008}. Polarized emission was initially detected by \cite{SeguraCox2015} using CARMA at 1.3 mm and at a resolution of $\sim0.35"$ ($\sim 50$AU). They found polarized emission along the minor axis with a fractional polarization of $\sim 2.5 \%$. A higher resolution, $\sim 0.09"$ ($\sim 13$AU), polarization map at 0.87~mm was obtained by \cite{Harris2018} using ALMA. It shows a polarization orientation along the minor axis and a polarization fraction of $\sim 3\%$ near the continuum peak, both consistent with those at 1.3~mm. The higher resolution reveals intriguing spatial variation in polarization fraction. It decreases outwards along the disk midplane to $\lesssim 1\%$ near the ends of the (projected) disk but increases rapidly to $\sim 10\%$ near the disk surface. As stressed by \cite{SeguraCox2015} and \cite{Harris2018}, in a nearly edge-on disk such as L1527, an orientation along the minor axis is consistent with the polarization from grains aligned by a toroidal magnetic field, at least in the optically thin limit. The outward decrease of polarization fraction along the disk midplane can be naturally explained as a result of the varying magnetic field orientation relative to the line of sight: the field at the center is perpendicular to the line of sight, which maximizes the polarization fraction, while that near the end is mostly along the line of sight. However, there is evidence that the disk is at least moderately optically thick, based on a comparison of the brightness temperature at 0.87~mm \citep{Harris2018} and the temperature distribution inferred from optically thick $^{13}$CO (2-1) and C$^{18}$O (2-1) lines \citep{vantHoff2018}. The higher optical depth towards the central position should reduce the polarization fraction more relative to those at larger radii, potentially making the interpretation involving grains aligned with a toroidal magnetic field inconsistent with the observed trend along the disk midplane.

An alternative interpretation, also discussed in \cite{Harris2018}, is that the grains are aligned by a vertical (or poloidal, rather than toroidal) magnetic field over most of the L1527 disk (except near the disk surface, see below). If the optical depth and temperature gradient along the line of sight is large enough, it would naturally produce a reversed polarization (parallel, rather than perpendicular, to the magnetic field direction in the sky plane). Evidence of a temperature gradient exists based on $^{13}$CO and C$^{18}$O observations which indicate higher temperature at a smaller radius \citep{vantHoff2018}. In this interpretation, the lower polarization fraction towards the ends of the disk on the midplane would be caused by a combination of a smaller optical depth, which is large enough to reverse the polarization but not large enough for the polarization fraction to reach its maximum value set by the temperature gradient $\psi$ near the $\tau'=1$ surface, and a smaller $\psi$, as illustrated in Figure \ref{fig:Tirr_obs}l and Figure \ref{fig:Tacc_obs}l. A difficulty with this vertical field scenario is that it would predict a polarization orientation parallel to the midplane near the disk surface where the dust is expected to be optically thin (Figure \ref{fig:Tirr_obs}). This is the opposite of what is observed, 
unless the surface is still optically thick. A potential solution to this problem is to have the predominantly vertical magnetic field in the bulk of the disk change into a predominantly toroidal configuration near the surface. Such a field variation could be produced, for example, in a lightly ionized disk, where the magnetic field is weakly coupled to the bulk of the disk material (and can thus remain vertical) except near the more ionized surface, where the magnetic field can be dragged by the gas rotation into a predominantly toroidal configuration, perhaps at the base of a disk wind \cite[see, e.g., Figure 2a and b of][]{Li1996}. In this interpretation, it would be natural to use the polarization fraction near the surface as an estimate of the intrinsic polarization fraction, i.e., $p_0\approx -10\%$, which would yield a temperature gradient towards the disk center (where the polarization is reversed in this interpretation, with an observed fraction of $p\approx 3\%$) of $\psi\approx -p/p_0\approx 0.3$ (according to Equation (\ref{eq:polpsi})). Whether this gradient quantitatively agrees with that inferred from optically thick CO lines \citep{vantHoff2018} or not remains to be determined. It will depend on the opacity of the dust grains relative to those of the CO (or other) lines used to infer the temperature gradient independently. This dependence may provide a rare handle on the grain opacity, a crucial quantity for determining the mass of solids in disks, the raw material for planetesimals and ultimately planets. Temperature gradient from polarization may also complement the method based on molecular line observations where the dust can be optically thick and lines may not be detectable. We will postpone a more detailed exploration of the connection between the temperature gradients inferred from dust polarization and molecular lines to a future investigation.

We should note that, in the above interpretation involving magnetically aligned grains, we have made the usual assumption that the rapidly spinning grains are aligned with their short axes along the magnetic field so that their (averaged) shapes are effectively oblate. It is possible to have effectively prolate (rather than oblate) shapes as well, as in the case of aerodynamic grain alignment, i.e., the Gold mechanism \citep{Gold1952, Lazarian1995}. In this case, the observed polarization pattern in L1527 would be consistent with a scenario where the (effectively) prolate grains are aligned in the toroidal direction in the optically thick part of the disk (where the polarization is reversed) and in the vertical direction in the optically thin part near the surface. The former could perhaps be produced by the streaming of dust grains relative to the gas in the azimuthal direction \citep{Yang2019, Kataoka2019}, whereas the latter by a wind expanding away from the disk. Compared to the scenario involving effectively oblate grains aligned with a vertical magnetic field in the bulk of the disk, described in the last paragraph, this scenario has the advantage that the effectively prolate grains are viewed more edge-on towards the center of the disk than towards both ends, making it easier to explain the observed decrease of polarization fraction with radius along the major axis. Whether the grains can move faster enough relative to the gas to produce a degree of alignment needed to explain the observed polarization level remains to be determined, however \citep{Andersson2015}. 

The disk of the Class 0 protostar HH 212 in Orion has an inclination angle of $\sim 86^\circ$, similar to that of L1527. Its spatially resolved polarization at 0.87~mm (ALMA Band 7) is oriented roughly along the minor axis, broadly similar to that for L1527 \citep{LeeCF2018}. HH 212's disk shows strong evidence that it is optically thick towards the midplane based on the prominent dark equatorial lane sandwiched between two brighter surface regions (\citealt{LeeCF2017_lane}; Z. Lin et al. in prep). Such a feature was also found in our edge-on models (Figures \ref{fig:Tirr_obs}l and \ref{fig:Tacc_obs}l). This makes polarization reversal likely for the disk, as already stressed by \cite{LeeCF2018}. In such a case, the observed polarization orientations along the minor axis are consistent with either effectively oblate grains that are poloidally aligned or effectively prolate grains that are toroidally aligned, as discussed earlier for L1527. As such, HH 212 is another good target for probing the disk temperature gradient through dust polarization, especially if the intrinsic polarization fraction $p_0$ can be estimated from future polarization observations that probe more optically thin regions, and for independently checking the inferred temperature gradient through line observations.

\begin{figure*}
    \centering
    \includegraphics[width=0.75\textwidth]{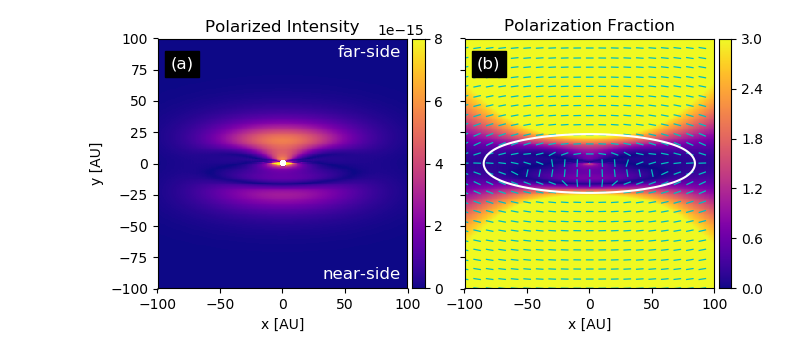}
    \caption{
            The polarized intensity and polarization fraction for the passive disk but with toroidally aligned prolate grains at an inclination of $75^{\circ}$. The panels are plotted in a similar manner as panels h and i presented in Figures \ref{fig:Tirr_obs} and \ref{fig:Tacc_obs}. 
        }
    \label{fig:prolate_i75}
\end{figure*}

The Class I disk, HH 111 VLA 1, has a lower inclination angle ($\sim 72^{\circ}$; \citealt{LeeCF2019}) than L1527 and HH 212, which is similar to the value chosen for our $75^{\circ}$ inclination case. Its spatially resolved polarization at 0.87~mm is predominantly along the minor axis in the near-side of the disk but along the major axis in the far-side. \cite{LeeCF2018} noted that this pattern is consistent with the polarization from effectively oblate grains aligned with a toroidal magnetic field in the near-side and with a poloidal (nearly vertical) field in the far-side if the disk is optically thin. However, in the scenario where the disk is optically thick, the polarization in the near-side is likely reversed whereas that in the far-side is not, because of the difference in temperature gradient along sight lines through the near- and far-side of the disk, as discussed earlier in Section \ref{ssec:passive_results} (see Figure \ref{fig:nearfar} for illustration). This provides a natural explanation of their orthogonal polarization orientations without invoking a $90^\circ$ change in the grain alignment axis. For example, effectively oblate grains aligned with a predominantly poloidal (vertical) magnetic field would produce polarization that is roughly parallel to the minor axis through the near-side via polarization reversal but parallel to the major axis in the far-side, as observed. Similarly, effectively prolate grains aligned in the toroidal direction (through, e.g., Gold mechanism) would produce an elliptical polarization orientation in the far-side (in the same direction as the optically thin case; \citealt{Yang2019}) but a radial orientation in the near-side through polarization reversal, again broadly consistent with the observed pattern. This can be seen in Figure \ref{fig:prolate_i75} where we have used the same density and temperature structure of the passive disk with inclination $75^{\circ}$, but toroidally aligned prolate (rather than oblate) grains.

The polarized intensity observed in the HH 111 disk is qualitatively similar to the highly inclination case shown in Figure \ref{fig:prolate_i75}a (see also Figures \ref{fig:Tirr_obs}h and \ref{fig:Tacc_obs}h). Specifically, the near-side has an elongated region of polarization intensity that becomes stronger towards the center. In transition to the far-side along the minor axis, the polarized intensity first decreases to zero and then increases again. Along the major axis, the polarization fraction decreases outwards because the toroidally aligned prolate grains are viewed more pole-on towards the disk edge, in agreement with the observed trend. 

To verify if temperature gradient is indeed at play, observations with higher sensitivity at the same wavelength is needed to detect polarization in the more optically thin regions of the near-side. As seen in Figure \ref{fig:prolate_i75}b (see also Figures \ref{fig:Tirr_obs}i and \ref{fig:Tacc_obs}i), there will no longer be a polarization reversal in such regions. Another way to verify the scenario is to observe polarization at longer wavelengths. The optically thick region should shrink in size towards the center, along with the region of reversed polarization, i.e., the outer region of the near-side with polarization along the minor axis at Band 7 would flip to the major axis direction. This  behavior is a telltale sign of aligned grains, particularly toroidally aligned prolate grains in this case. As discussed in Section \ref{sec:analytical}, the polarization fraction in the region with polarization reversal serves as a lower limit to the intrinsic polarization of the grains. Thus, prolate grains, if they are indeed responsible for the polarization observed in the HH 111 disk, should have an intrinsic polarization of at least the value observed in the region of polarization reversal ($\sim 1 \%$). 


%
%
%
%
%
%





We note that although the discussion in this section has been focused mostly on the relatively young (Class 0 and early Class I) disks that are either intrinsically bright (such as NGC IRAS4A1) or highly inclined (such as HH 212), the temperature gradient effect on the dust polarization can also be significant in the fainter, more evolved (Class II) disks at relatively small radii where the optical depth can be high and viscous heating is expected to be important \citep[e.g.][]{Bitsch2015}. However, to detect this effect, a higher resolution would be required, making it harder to do with the current observational capabilities.

\section{Summary and Conclusions} \label{sec:Conclusions}

In this paper, we have investigated the effects of temperature gradient on the polarized emission from aligned grains, with emphasis on disks around young stellar objects. Our main conclusions are as follows:


(1) The disk polarization from aligned grains is affected by the temperature gradient most strongly along sight lines that are optically thick. This is because the light polarized along the short axes of the aligned grains (projected on the sky plane) has a lower opacity than that polarized along the long axes and thus comes mostly from a region deeper into the disk if the sight line is optically thick, as illustrated in Figure \ref{fig:schem}. As the optical depth of a sight line increases, its polarization fraction quickly approaches zero in the case of zero temperature gradient. With a decreasing temperature along the line of sight, the polarization fraction asymptotes to a non-zero value, with the same polarization direction as that in the optically thin limit. With an increasing temperature along the line of sight, the fraction approaches a constant value as well, but with a polarization direction along the (projected) short (rather than long) axes of the aligned grains, which is orthogonal to the optically thin case. This ``reversed" polarization along optically thick sight lines with increasing temperature away from the observer is a telltale sign of the polarization from aligned grains.  


(2) The asymptotic value of the polarization fraction $p$ in the optically thick limit can be estimated analytically in a simple 1D slab model with a constant temperature gradient $dT/d\tau'$ relative to the optical depth $\tau'$. It depends only on the intrinsic polarization fraction $p_0$ (the polarization fraction at the same wavelength and along the same sight line in the optically thin limit) and the logarithmic temperature gradient near the $\tau'=1$ surface along the line of sight in the Rayleigh Jeans limit, through $p\approx -p_0 (d\ln T/d\tau'\vert_{\tau'=1})$ (Equation (\ref{eq:polpsi})). This formula provides a way to estimate the temperature gradient along an optically thick line of sight through the disk from the normalized polarization fraction $-p/p_0$, which can be observationally measured in principle.


(3) The above approximate formula (Equation (\ref{eq:polpsi})) for inferring temperature gradient from dust polarization is validated using a model of stellar irradiated disks with and without accretion heating. In both cases, it holds to a factor of 2 even in the presence of a strong variation in temperature gradient and a significant deviation from the Rayleigh Jeans limit. As expected, we find that the polarization is reversed along optically thick sight lines towards an edge-on disk (with an inclination angle of $i=90^\circ$) as a result of the temperature increasing radially towards the central stellar object (see Figures \ref{fig:Tirr_obs}l and \ref{fig:Tacc_obs}l). For face-on disks ($i=0^\circ$), polarization reversal can also happen, but only for actively accreting disks where the temperature increases towards the midplane (Figure \ref{fig:Tacc_obs}c); in other words, dust polarization can in principle provide a powerful probe of disk accretion. Most intriguing is the $75^{\circ}$ inclination case, where the near-side of the disk is viewed more edge-on than the far-side and the traced temperature gradients are different (see Figure \ref{fig:nearfar} for illustration). As a result, the polarization is more likely to reverse in the near-side than the far-side, producing a polarization orientation in the near-side that is orthogonal to that on the far-side (see Figures \ref{fig:Tirr_obs}i and \ref{fig:Tacc_obs}i). The $90^\circ$ flip between the near- and far-side is another telltale sign of the polarization produced by aligned grains.  


(4) The best targets to look for the temperature gradient effects on dust polarization are the disks that are optically thick, such as those that have a high surface density or are nearly edge-on. One intriguing case is the relatively face-on, very bright disk-like structure in the deeply embedded Class 0 object NGC1333 IRAS4A1, whose polarization orientations at the very optically thick (sub)millimeter wavelengths are orthogonal to those at the more optically thin centimeter wavelengths, which is suggestive of a polarization reversal at the shorter wavelengths produced by an accretion-heated midplane (see Figure \ref{fig:Tacc_obs}c). Another is the highly inclined (but not quite edge-on) disk of the Class I object HH 111, whose polarization along the edge-on sight lines through the near-side of the disk appears reversed relative to that along the more face-on sight lines through the far-side; the former probes the temperature gradient predominantly in the radial direction whereas the latter in the vertical direction. The polarization fraction observed in the region of reversed polarization allows for an estimate of the temperature gradient with respect to the optical depth $\tau'$ from Equation (\ref{eq:polpsi}), if the intrinsic polarization fraction $p_0$ can be estimated, from polarization either along more optically thin sight lines or at longer wavelengths. High resolution and sensitivity polarization observations at widely separated wavebands are desirable for this task. 

(5) Our technique of inferring the disk temperature structure from dust polarization measures the temperature gradient relative to the dust optical depth, and complements those using other methods, particularly molecular lines. Indeed, when combined with independent measurements of the temperature structure, such as through molecular lines, dust polarization may provide a rare handle on the grain opacity, and thus the mass of the solids, the raw material for forming planetesimals and ultimately planets. 

\section*{Acknowledgements}

We thank the referee for constructive comments. This research made use of radmc3dPy \footnote{\url{https://www.ast.cam.ac.uk/~juhasz/radmc3dPyDoc/index.html}}, a python package, for front end use of RADMC-3D. ZYDL acknowledges an ALMA SOS award from National Radio Astronomy Observatory. ZYL is supported in part by NASA 80NSSC18K1095 and NSF AST-1716259, 1815784, and 1910106. This paper makes use of the following ALMA data: ADS/JAO.ALMA\#2015.1.00546.S, ADS/JAO.ALMA\#2016.1.01089.S, ADS/JAO.ALMA\#2015.1.01112.S, ADS/JAO.ALMA\#2013.1.00291.S, ADS/JAO.ALMA\#2016.1.01186.S, ADS/JAO.ALMA\#2015.1.00084.S, and ADS/JAO.ALMA\#2015.1.00037.S. ALMA is a partnership of ESO (representing its member states), NSF (USA) and NINS (Japan), together with NRC (Canada), MOST and ASIAA (Taiwan), and KASI (Republic of Korea), in cooperation with the Republic of Chile. The Joint ALMA Observatory is operated by ESO, AUI/NRAO and NAOJ. The National Radio Astronomy Observatory is a facility of the National Science Foundation operated under cooperative agreement by Associated Universities, Inc.


\bibliographystyle{mnras}
\bibliography{paper} 

\begin{thebibliography}{}
\makeatletter
\relax
\def\mn@urlcharsother{\let\do\@makeother \do\$\do\&\do\#\do\^\do\_\do\%\do\~}
\def\mn@doi{\begingroup\mn@urlcharsother \@ifnextchar [ {\mn@doi@}
  {\mn@doi@[]}}
\def\mn@doi@[#1]#2{\def\@tempa{#1}\ifx\@tempa\@empty \href
  {http://dx.doi.org/#2} {doi:#2}\else \href {http://dx.doi.org/#2} {#1}\fi
  \endgroup}
\def\mn@eprint#1#2{\mn@eprint@#1:#2::\@nil}
\def\mn@eprint@arXiv#1{\href {http://arxiv.org/abs/#1} {{\tt arXiv:#1}}}
\def\mn@eprint@dblp#1{\href {http://dblp.uni-trier.de/rec/bibtex/#1.xml}
  {dblp:#1}}
\def\mn@eprint@#1:#2:#3:#4\@nil{\def\@tempa {#1}\def\@tempb {#2}\def\@tempc
  {#3}\ifx \@tempc \@empty \let \@tempc \@tempb \let \@tempb \@tempa \fi \ifx
  \@tempb \@empty \def\@tempb {arXiv}\fi \@ifundefined
  {mn@eprint@\@tempb}{\@tempb:\@tempc}{\expandafter \expandafter \csname
  mn@eprint@\@tempb\endcsname \expandafter{\@tempc}}}

\bibitem[\protect\citeauthoryear{{Agurto-Gangas} et~al.,}{{Agurto-Gangas}
  et~al.}{2019}]{Agurto2019}
{Agurto-Gangas} C.,  et~al., 2019, \mn@doi [\aap]
  {10.1051/0004-6361/201833666}, \href
  {https://ui.adsabs.harvard.edu/abs/2019A&A...623A.147A} {623, A147}

\bibitem[\protect\citeauthoryear{{Alves} et~al.,}{{Alves}
  et~al.}{2018}]{Alves2018}
{Alves} F.~O.,  et~al., 2018, \mn@doi [\aap] {10.1051/0004-6361/201832935},
  \href {http://adsabs.harvard.edu/abs/2018A%26A...616A..56A} {616, A56}

\bibitem[\protect\citeauthoryear{{Andersson}, {Lazarian}  \&
  {Vaillancourt}}{{Andersson} et~al.}{2015}]{Andersson2015}
{Andersson} B.-G.,  {Lazarian} A.,   {Vaillancourt} J.~E.,  2015, \mn@doi
  [\araa] {10.1146/annurev-astro-082214-122414}, \href
  {http://adsabs.harvard.edu/abs/2015ARA%26A..53..501A} {53, 501}

\bibitem[\protect\citeauthoryear{{Andrews}, {Wilner}, {Hughes}, {Qi}  \&
  {Dullemond}}{{Andrews} et~al.}{2010}]{Andrews2010}
{Andrews} S.~M.,  {Wilner} D.~J.,  {Hughes} A.~M.,  {Qi} C.,   {Dullemond}
  C.~P.,  2010, \mn@doi [\apj] {10.1088/0004-637X/723/2/1241}, \href
  {https://ui.adsabs.harvard.edu/abs/2010ApJ...723.1241A} {723, 1241}

\bibitem[\protect\citeauthoryear{{Bacciotti} et~al.,}{{Bacciotti}
  et~al.}{2018}]{Bacciotti2018}
{Bacciotti} F.,  et~al., 2018, \mn@doi [\apjl] {10.3847/2041-8213/aadf87},
  \href {http://adsabs.harvard.edu/abs/2018ApJ...865L..12B} {865, L12}

\bibitem[\protect\citeauthoryear{{Balbus} \& {Hawley}}{{Balbus} \&
  {Hawley}}{1991}]{Balbus1991}
{Balbus} S.~A.,  {Hawley} J.~F.,  1991, \mn@doi [\apj] {10.1086/170270}, \href
  {http://adsabs.harvard.edu/abs/1991ApJ...376..214B} {376, 214}

\bibitem[\protect\citeauthoryear{{Beckwith}, {Sargent}, {Chini}  \&
  {Guesten}}{{Beckwith} et~al.}{1990}]{Beckwith1990}
{Beckwith} S. V.~W.,  {Sargent} A.~I.,  {Chini} R.~S.,   {Guesten} R.,  1990,
  \mn@doi [\aj] {10.1086/115385}, \href
  {https://ui.adsabs.harvard.edu/abs/1990AJ.....99..924B} {99, 924}

\bibitem[\protect\citeauthoryear{{Bitsch}, {Johansen}, {Lambrechts}  \&
  {Morbidelli}}{{Bitsch} et~al.}{2015}]{Bitsch2015}
{Bitsch} B.,  {Johansen} A.,  {Lambrechts} M.,   {Morbidelli} A.,  2015,
  \mn@doi [\aap] {10.1051/0004-6361/201424964}, \href
  {https://ui.adsabs.harvard.edu/abs/2015A&A...575A..28B} {575, A28}

\bibitem[\protect\citeauthoryear{{Blandford} \& {Payne}}{{Blandford} \&
  {Payne}}{1982}]{Blandford1982}
{Blandford} R.~D.,  {Payne} D.~G.,  1982, \mn@doi [\mnras]
  {10.1093/mnras/199.4.883}, \href
  {http://adsabs.harvard.edu/abs/1982MNRAS.199..883B} {199, 883}

\bibitem[\protect\citeauthoryear{{Bohren} \& {Huffman}}{{Bohren} \&
  {Huffman}}{1983}]{Bohren1983}
{Bohren} C.~F.,  {Huffman} D.~R.,  1983, {Absorption and scattering of light by
  small particles}

\bibitem[\protect\citeauthoryear{{Cleeves}, {{\"O}berg}, {Wilner}, {Huang},
  {Loomis}, {Andrews}  \& {Czekala}}{{Cleeves} et~al.}{2016}]{Cleeves2016}
{Cleeves} L.~I.,  {{\"O}berg} K.~I.,  {Wilner} D.~J.,  {Huang} J.,  {Loomis}
  R.~A.,  {Andrews} S.~M.,   {Czekala} I.,  2016, \mn@doi [\apj]
  {10.3847/0004-637X/832/2/110}, \href
  {http://adsabs.harvard.edu/abs/2016ApJ...832..110C} {832, 110}

\bibitem[\protect\citeauthoryear{{Cox} et~al.,}{{Cox} et~al.}{2015}]{Cox2015}
{Cox} E.~G.,  et~al., 2015, \mn@doi [\apjl] {10.1088/2041-8205/814/2/L28},
  \href {https://ui.adsabs.harvard.edu/abs/2015ApJ...814L..28C} {814, L28}

\bibitem[\protect\citeauthoryear{{D'Alessio}, {Cant{\"o}}, {Calvet}  \&
  {Lizano}}{{D'Alessio} et~al.}{1998}]{DAlessio1998}
{D'Alessio} P.,  {Cant{\"o}} J.,  {Calvet} N.,   {Lizano} S.,  1998, \mn@doi
  [\apj] {10.1086/305702}, \href
  {http://adsabs.harvard.edu/abs/1998ApJ...500..411D} {500, 411}

\bibitem[\protect\citeauthoryear{{Dent}, {Pinte}, {Cortes}, {M{\'e}nard},
  {Hales}, {Fomalont}  \& {de Gregorio-Monsalvo}}{{Dent}
  et~al.}{2019}]{Dent2019}
{Dent} W.~R.~F.,  {Pinte} C.,  {Cortes} P.~C.,  {M{\'e}nard} F.,  {Hales} A.,
  {Fomalont} E.,   {de Gregorio-Monsalvo} I.,  2019, \mn@doi [\mnras]
  {10.1093/mnrasl/sly181}, \href
  {https://ui.adsabs.harvard.edu/abs/2019MNRAS.482L..29D} {482, L29}

\bibitem[\protect\citeauthoryear{{Draine} \& {Weingartner}}{{Draine} \&
  {Weingartner}}{1997}]{Draine1997}
{Draine} B.~T.,  {Weingartner} J.~C.,  1997, \mn@doi [\apj] {10.1086/304008},
  \href {http://adsabs.harvard.edu/abs/1997ApJ...480..633D} {480, 633}

\bibitem[\protect\citeauthoryear{{Dullemond}}{{Dullemond}}{2002}]{Dullemond2002_2D}
{Dullemond} C.~P.,  2002, \mn@doi [\aap] {10.1051/0004-6361:20021300}, \href
  {http://adsabs.harvard.edu/abs/2002A%26A...395..853D} {395, 853}

\bibitem[\protect\citeauthoryear{{Dullemond}, {van Zadelhoff}  \&
  {Natta}}{{Dullemond} et~al.}{2002}]{Dullemond2002_varied}
{Dullemond} C.~P.,  {van Zadelhoff} G.~J.,   {Natta} A.,  2002, \mn@doi [\aap]
  {10.1051/0004-6361:20020608}, \href
  {http://adsabs.harvard.edu/abs/2002A%26A...389..464D} {389, 464}

\bibitem[\protect\citeauthoryear{{Girart}, {Rao}  \& {Marrone}}{{Girart}
  et~al.}{2006}]{Girart2006}
{Girart} J.~M.,  {Rao} R.,   {Marrone} D.~P.,  2006, \mn@doi [Science]
  {10.1126/science.1129093}, \href
  {http://adsabs.harvard.edu/abs/2006Sci...313..812G} {313, 812}

\bibitem[\protect\citeauthoryear{{Girart} et~al.,}{{Girart}
  et~al.}{2018}]{Girart2018}
{Girart} J.~M.,  et~al., 2018, \mn@doi [\apjl] {10.3847/2041-8213/aab76b},
  \href {https://ui.adsabs.harvard.edu/abs/2018ApJ...856L..27G} {856, L27}

\bibitem[\protect\citeauthoryear{{Gold}}{{Gold}}{1952}]{Gold1952}
{Gold} T.,  1952, \mn@doi [\mnras] {10.1093/mnras/112.2.215}, \href
  {http://adsabs.harvard.edu/abs/1952MNRAS.112..215G} {112, 215}

\bibitem[\protect\citeauthoryear{{Harris} et~al.,}{{Harris}
  et~al.}{2018}]{Harris2018}
{Harris} R.~J.,  et~al., 2018, \mn@doi [\apj] {10.3847/1538-4357/aac6ec}, \href
  {https://ui.adsabs.harvard.edu/abs/2018ApJ...861...91H} {861, 91}

\bibitem[\protect\citeauthoryear{{Hartmann}, {D'Alessio}, {Calvet}  \&
  {Muzerolle}}{{Hartmann} et~al.}{2006}]{Hartmann2006}
{Hartmann} L.,  {D'Alessio} P.,  {Calvet} N.,   {Muzerolle} J.,  2006, \mn@doi
  [\apj] {10.1086/505788}, \href
  {https://ui.adsabs.harvard.edu/abs/2006ApJ...648..484H} {648, 484}

\bibitem[\protect\citeauthoryear{{Henning} \& {Stognienko}}{{Henning} \&
  {Stognienko}}{1996}]{Henning1996}
{Henning} T.,  {Stognienko} R.,  1996, \aap, \href
  {https://ui.adsabs.harvard.edu/abs/1996A&A...311..291H} {311, 291}

\bibitem[\protect\citeauthoryear{{Hildebrand}, {Davidson}, {Dotson}, {Dowell},
  {Novak}  \& {Vaillancourt}}{{Hildebrand} et~al.}{2000}]{Hildebrand2000}
{Hildebrand} R.~H.,  {Davidson} J.~A.,  {Dotson} J.~L.,  {Dowell} C.~D.,
  {Novak} G.,   {Vaillancourt} J.~E.,  2000, \mn@doi [\pasp] {10.1086/316613},
  \href {http://adsabs.harvard.edu/abs/2000PASP..112.1215H} {112, 1215}

\bibitem[\protect\citeauthoryear{{Hull} et~al.,}{{Hull}
  et~al.}{2018}]{Hull2018}
{Hull} C.~L.~H.,  et~al., 2018, \mn@doi [\apj] {10.3847/1538-4357/aabfeb},
  \href {http://adsabs.harvard.edu/abs/2018ApJ...860...82H} {860, 82}

\bibitem[\protect\citeauthoryear{{Kataoka} et~al.,}{{Kataoka}
  et~al.}{2015}]{Kataoka2015}
{Kataoka} A.,  et~al., 2015, \mn@doi [\apj] {10.1088/0004-637X/809/1/78}, \href
  {http://adsabs.harvard.edu/abs/2015ApJ...809...78K} {809, 78}

\bibitem[\protect\citeauthoryear{{Kataoka} et~al.,}{{Kataoka}
  et~al.}{2016}]{Kataoka2016}
{Kataoka} A.,  et~al., 2016, \mn@doi [\apjl] {10.3847/2041-8205/831/2/L12},
  \href {http://adsabs.harvard.edu/abs/2016ApJ...831L..12K} {831, L12}

\bibitem[\protect\citeauthoryear{{Kataoka}, {Tsukagoshi}, {Pohl}, {Muto},
  {Nagai}, {Stephens}, {Tomisaka}  \& {Momose}}{{Kataoka}
  et~al.}{2017}]{Kataoka2017}
{Kataoka} A.,  {Tsukagoshi} T.,  {Pohl} A.,  {Muto} T.,  {Nagai} H.,
  {Stephens} I.~W.,  {Tomisaka} K.,   {Momose} M.,  2017, \mn@doi [\apjl]
  {10.3847/2041-8213/aa7e33}, \href
  {http://adsabs.harvard.edu/abs/2017ApJ...844L...5K} {844, L5}

\bibitem[\protect\citeauthoryear{{Kataoka}, {Okuzumi}  \& {Tazaki}}{{Kataoka}
  et~al.}{2019}]{Kataoka2019}
{Kataoka} A.,  {Okuzumi} S.,   {Tazaki} R.,  2019, \mn@doi [\apjl]
  {10.3847/2041-8213/ab0c9a}, \href
  {https://ui.adsabs.harvard.edu/abs/2019ApJ...874L...6K} {874, L6}

\bibitem[\protect\citeauthoryear{{Ko}, {Liu}, {Lai}, {Ching}, {Rao}  \&
  {Girart}}{{Ko} et~al.}{2020}]{Ko2020}
{Ko} C.-L.,  {Liu} H.~B.,  {Lai} S.-P.,  {Ching} T.-C.,  {Rao} R.,   {Girart}
  J.~M.,  2020, \mn@doi [\apj] {10.3847/1538-4357/ab5e79}, \href
  {https://ui.adsabs.harvard.edu/abs/2020ApJ...889..172K} {889, 172}

\bibitem[\protect\citeauthoryear{{Lazarian}}{{Lazarian}}{1995}]{Lazarian1995}
{Lazarian} A.,  1995, \mn@doi [\apj] {10.1086/176252}, \href
  {http://adsabs.harvard.edu/abs/1995ApJ...451..660L} {451, 660}

\bibitem[\protect\citeauthoryear{{Lazarian} \& {Hoang}}{{Lazarian} \&
  {Hoang}}{2007a}]{Lazarian2007}
{Lazarian} A.,  {Hoang} T.,  2007a, \mn@doi [\mnras]
  {10.1111/j.1365-2966.2007.11817.x}, \href
  {http://adsabs.harvard.edu/abs/2007MNRAS.378..910L} {378, 910}

\bibitem[\protect\citeauthoryear{{Lazarian} \& {Hoang}}{{Lazarian} \&
  {Hoang}}{2007b}]{Lazarian2007_MAT}
{Lazarian} A.,  {Hoang} T.,  2007b, \mn@doi [\apjl] {10.1086/523849}, \href
  {http://adsabs.harvard.edu/abs/2007ApJ...669L..77L} {669, L77}

\bibitem[\protect\citeauthoryear{{Lee} \& {Draine}}{{Lee} \&
  {Draine}}{1985}]{LeeDraine1985}
{Lee} H.~M.,  {Draine} B.~T.,  1985, \mn@doi [\apj] {10.1086/162974}, \href
  {http://adsabs.harvard.edu/abs/1985ApJ...290..211L} {290, 211}

\bibitem[\protect\citeauthoryear{{Lee}, {Li}, {Ho}, {Hirano}, {Zhang}  \&
  {Shang}}{{Lee} et~al.}{2017}]{LeeCF2017_lane}
{Lee} C.-F.,  {Li} Z.-Y.,  {Ho} P. T.~P.,  {Hirano} N.,  {Zhang} Q.,   {Shang}
  H.,  2017, \mn@doi [Science Advances] {10.1126/sciadv.1602935}, \href
  {https://ui.adsabs.harvard.edu/abs/2017SciA....3E2935L} {3, e1602935}

\bibitem[\protect\citeauthoryear{{Lee}, {Li}, {Ching}, {Lai}  \& {Yang}}{{Lee}
  et~al.}{2018}]{LeeCF2018}
{Lee} C.-F.,  {Li} Z.-Y.,  {Ching} T.-C.,  {Lai} S.-P.,   {Yang} H.,  2018,
  \mn@doi [\apj] {10.3847/1538-4357/aaa769}, \href
  {http://adsabs.harvard.edu/abs/2018ApJ...854...56L} {854, 56}

\bibitem[\protect\citeauthoryear{{Lee}, {Li}  \& {Turner}}{{Lee}
  et~al.}{2019}]{LeeCF2019}
{Lee} C.-F.,  {Li} Z.-Y.,   {Turner} N.~J.,  2019, \mn@doi [Nature Astronomy]
  {10.1038/s41550-019-0905-x}, \href
  {https://ui.adsabs.harvard.edu/abs/2019NatAs.tmp..466L} {p.~466}

\bibitem[\protect\citeauthoryear{{Li}}{{Li}}{1996}]{Li1996}
{Li} Z.-Y.,  1996, \mn@doi [\apj] {10.1086/177469}, \href
  {https://ui.adsabs.harvard.edu/abs/1996ApJ...465..855L} {465, 855}

\bibitem[\protect\citeauthoryear{{Liu} et~al.,}{{Liu} et~al.}{2016}]{Liu2016}
{Liu} H.~B.,  et~al., 2016, \mn@doi [\apj] {10.3847/0004-637X/821/1/41}, \href
  {https://ui.adsabs.harvard.edu/abs/2016ApJ...821...41L} {821, 41}

\bibitem[\protect\citeauthoryear{{Liu}, {Hasegawa}, {Ching}, {Lai}, {Hirano}
  \& {Rao}}{{Liu} et~al.}{2018}]{Liu2018}
{Liu} H.~B.,  {Hasegawa} Y.,  {Ching} T.-C.,  {Lai} S.-P.,  {Hirano} N.,
  {Rao} R.,  2018, \mn@doi [\aap] {10.1051/0004-6361/201832699}, \href
  {https://ui.adsabs.harvard.edu/abs/2018A&A...617A...3L} {617, A3}

\bibitem[\protect\citeauthoryear{{Looney}, {Mundy}  \& {Welch}}{{Looney}
  et~al.}{2000}]{Looney2000}
{Looney} L.~W.,  {Mundy} L.~G.,   {Welch} W.~J.,  2000, \mn@doi [\apj]
  {10.1086/308239}, \href
  {https://ui.adsabs.harvard.edu/abs/2000ApJ...529..477L} {529, 477}

\bibitem[\protect\citeauthoryear{{Lynden-Bell} \& {Pringle}}{{Lynden-Bell} \&
  {Pringle}}{1974}]{LyndenBell1974}
{Lynden-Bell} D.,  {Pringle} J.~E.,  1974, \mn@doi [\mnras]
  {10.1093/mnras/168.3.603}, \href
  {http://adsabs.harvard.edu/abs/1974MNRAS.168..603L} {168, 603}

\bibitem[\protect\citeauthoryear{{Morbidelli} \& {Raymond}}{{Morbidelli} \&
  {Raymond}}{2016}]{Morbidelli2016}
{Morbidelli} A.,  {Raymond} S.~N.,  2016, \mn@doi [Journal of Geophysical
  Research (Planets)] {10.1002/2016JE005088}, \href
  {http://adsabs.harvard.edu/abs/2016JGRE..121.1962M} {121, 1962}

\bibitem[\protect\citeauthoryear{{Mulders}, {Pascucci}, {Manara}, {Testi},
  {Herczeg}, {Henning}, {Mohanty}  \& {Lodato}}{{Mulders}
  et~al.}{2017}]{Mulders2017}
{Mulders} G.~D.,  {Pascucci} I.,  {Manara} C.~F.,  {Testi} L.,  {Herczeg}
  G.~J.,  {Henning} T.,  {Mohanty} S.,   {Lodato} G.,  2017, \mn@doi [\apj]
  {10.3847/1538-4357/aa8906}, \href
  {https://ui.adsabs.harvard.edu/abs/2017ApJ...847...31M} {847, 31}

\bibitem[\protect\citeauthoryear{{Natta}, {Testi}  \& {Randich}}{{Natta}
  et~al.}{2006}]{Natta2006}
{Natta} A.,  {Testi} L.,   {Randich} S.,  2006, \mn@doi [Astronomy and
  Astrophysics] {10.1051/0004-6361:20054706}, \href
  {https://ui.adsabs.harvard.edu/abs/2006A&A...452..245N} {452, 245}

\bibitem[\protect\citeauthoryear{{Pineda} et~al.,}{{Pineda}
  et~al.}{2012}]{Pineda2012}
{Pineda} J.~E.,  et~al., 2012, \mn@doi [\aap] {10.1051/0004-6361/201219589},
  \href {https://ui.adsabs.harvard.edu/abs/2012A&A...544L...7P} {544, L7}

\bibitem[\protect\citeauthoryear{{Pinte} et~al.,}{{Pinte}
  et~al.}{2018}]{Pinte2018}
{Pinte} C.,  et~al., 2018, \mn@doi [\aap] {10.1051/0004-6361/201731377}, \href
  {http://adsabs.harvard.edu/abs/2018A%26A...609A..47P} {609, A47}

\bibitem[\protect\citeauthoryear{{Planck Collaboration} et~al.,}{{Planck
  Collaboration} et~al.}{2016}]{PlanckCollaboration2016}
{Planck Collaboration} et~al., 2016, \mn@doi [\aap]
  {10.1051/0004-6361/201525896}, \href
  {http://adsabs.harvard.edu/abs/2016A%26A...586A.138P} {586, A138}

\bibitem[\protect\citeauthoryear{{Pollack}, {Hollenbach}, {Beckwith},
  {Simonelli}, {Roush}  \& {Fong}}{{Pollack} et~al.}{1994}]{Pollack1994}
{Pollack} J.~B.,  {Hollenbach} D.,  {Beckwith} S.,  {Simonelli} D.~P.,  {Roush}
  T.,   {Fong} W.,  1994, \mn@doi [\apj] {10.1086/173677}, \href
  {http://adsabs.harvard.edu/abs/1994ApJ...421..615P} {421, 615}

\bibitem[\protect\citeauthoryear{{Rao}, {Girart}, {Lai}  \& {Marrone}}{{Rao}
  et~al.}{2014}]{Rao2014}
{Rao} R.,  {Girart} J.~M.,  {Lai} S.-P.,   {Marrone} D.~P.,  2014, \mn@doi
  [\apjl] {10.1088/2041-8205/780/1/L6}, \href
  {https://ui.adsabs.harvard.edu/abs/2014ApJ...780L...6R} {780, L6}

\bibitem[\protect\citeauthoryear{{Rosenfeld}, {Andrews}, {Hughes}, {Wilner}  \&
  {Qi}}{{Rosenfeld} et~al.}{2013}]{Rosenfeld2013}
{Rosenfeld} K.~A.,  {Andrews} S.~M.,  {Hughes} A.~M.,  {Wilner} D.~J.,   {Qi}
  C.,  2013, \mn@doi [\apj] {10.1088/0004-637X/774/1/16}, \href
  {http://adsabs.harvard.edu/abs/2013ApJ...774...16R} {774, 16}

\bibitem[\protect\citeauthoryear{{Sadavoy} et~al.,}{{Sadavoy}
  et~al.}{2018}]{Sadavoy2018}
{Sadavoy} S.~I.,  et~al., 2018, \mn@doi [\apj] {10.3847/1538-4357/aaef81},
  \href {https://ui.adsabs.harvard.edu/abs/2018ApJ...869..115S} {869, 115}

\bibitem[\protect\citeauthoryear{{Sahu}, {Liu}, {Su}, {Li}, {Lee}, {Hirano}  \&
  {Takakuwa}}{{Sahu} et~al.}{2019}]{Sahu2019}
{Sahu} D.,  {Liu} S.-Y.,  {Su} Y.-N.,  {Li} Z.-Y.,  {Lee} C.-F.,  {Hirano} N.,
   {Takakuwa} S.,  2019, \mn@doi [\apj] {10.3847/1538-4357/aaffda}, \href
  {https://ui.adsabs.harvard.edu/abs/2019ApJ...872..196S} {872, 196}

\bibitem[\protect\citeauthoryear{{Segura-Cox}, {Looney}, {Stephens},
  {Fern{\'a}ndez-L{\'o}pez}, {Kwon}, {Tobin}, {Li}  \& {Crutcher}}{{Segura-Cox}
  et~al.}{2015}]{SeguraCox2015}
{Segura-Cox} D.~M.,  {Looney} L.~W.,  {Stephens} I.~W.,
  {Fern{\'a}ndez-L{\'o}pez} M.,  {Kwon} W.,  {Tobin} J.~J.,  {Li} Z.-Y.,
  {Crutcher} R.,  2015, \mn@doi [\apjl] {10.1088/2041-8205/798/1/L2}, \href
  {https://ui.adsabs.harvard.edu/abs/2015ApJ...798L...2S} {798, L2}

\bibitem[\protect\citeauthoryear{{Segura-Cox} et~al.,}{{Segura-Cox}
  et~al.}{2018}]{SeguraCox2018}
{Segura-Cox} D.~M.,  et~al., 2018, \mn@doi [\apj] {10.3847/1538-4357/aaddf3},
  \href {https://ui.adsabs.harvard.edu/abs/2018ApJ...866..161S} {866, 161}

\bibitem[\protect\citeauthoryear{{Shakura} \& {Sunyaev}}{{Shakura} \&
  {Sunyaev}}{1973}]{Shakura1973}
{Shakura} N.~I.,  {Sunyaev} R.~A.,  1973, \aap, \href
  {https://ui.adsabs.harvard.edu/abs/1973A&A....24..337S} {500, 33}

\bibitem[\protect\citeauthoryear{{Stephens} et~al.,}{{Stephens}
  et~al.}{2013}]{Stephens2013}
{Stephens} I.~W.,  et~al., 2013, \mn@doi [The Astrophysical Journal]
  {10.1088/2041-8205/769/1/L15}, \href
  {https://ui.adsabs.harvard.edu/abs/2013ApJ...769L..15S} {769, L15}

\bibitem[\protect\citeauthoryear{{Stephens} et~al.,}{{Stephens}
  et~al.}{2014}]{Stephens2014}
{Stephens} I.~W.,  et~al., 2014, \mn@doi [\nat] {10.1038/nature13850}, \href
  {https://ui.adsabs.harvard.edu/abs/2014Natur.514..597S} {514, 597}

\bibitem[\protect\citeauthoryear{{Stephens} et~al.,}{{Stephens}
  et~al.}{2017}]{Stephens2017}
{Stephens} I.~W.,  et~al., 2017, \mn@doi [\apj] {10.3847/1538-4357/aa998b},
  \href {http://adsabs.harvard.edu/abs/2017ApJ...851...55S} {851, 55}

\bibitem[\protect\citeauthoryear{{Stephens} et~al.,}{{Stephens}
  et~al.}{2019}]{Stephens2019_whitepaper}
{Stephens} I.,  et~al., 2019, Bulletin of the American Astronomical Society,
  \href {https://ui.adsabs.harvard.edu/abs/2019BAAS...51c.246S} {51, 246}

\bibitem[\protect\citeauthoryear{{Su}, {Liu}, {Li}, {Lee}, {Hirano}, {Takakuwa}
   \& {Hsieh}}{{Su} et~al.}{2019}]{Su2019}
{Su} Y.-N.,  {Liu} S.-Y.,  {Li} Z.-Y.,  {Lee} C.-F.,  {Hirano} N.,  {Takakuwa}
  S.,   {Hsieh} I.-T.,  2019, \mn@doi [\apj] {10.3847/1538-4357/ab4818}, \href
  {https://iopscience.iop.org/article/10.3847/1538-4357/ab4818} {885, 98}

\bibitem[\protect\citeauthoryear{{Tazaki}, {Lazarian}  \& {Nomura}}{{Tazaki}
  et~al.}{2017}]{Tazaki2017}
{Tazaki} R.,  {Lazarian} A.,   {Nomura} H.,  2017, \mn@doi [\apj]
  {10.3847/1538-4357/839/1/56}, \href
  {http://adsabs.harvard.edu/abs/2017ApJ...839...56T} {839, 56}

\bibitem[\protect\citeauthoryear{{Tobin}, {Hartmann}, {Calvet}  \&
  {D'Alessio}}{{Tobin} et~al.}{2008}]{Tobin2008}
{Tobin} J.~J.,  {Hartmann} L.,  {Calvet} N.,   {D'Alessio} P.,  2008, \mn@doi
  [\apj] {10.1086/587683}, \href
  {https://ui.adsabs.harvard.edu/abs/2008ApJ...679.1364T} {679, 1364}

\bibitem[\protect\citeauthoryear{{Woitke} et~al.,}{{Woitke}
  et~al.}{2016}]{Woitke2016}
{Woitke} P.,  et~al., 2016, \mn@doi [\aap] {10.1051/0004-6361/201526538}, \href
  {https://ui.adsabs.harvard.edu/abs/2016A&A...586A.103W} {586, A103}

\bibitem[\protect\citeauthoryear{{Yang}, {Li}, {Looney}  \& {Stephens}}{{Yang}
  et~al.}{2016}]{Yang2016_HLTau}
{Yang} H.,  {Li} Z.-Y.,  {Looney} L.,   {Stephens} I.,  2016, \mn@doi [\mnras]
  {10.1093/mnras/stv2633}, \href
  {http://adsabs.harvard.edu/abs/2016MNRAS.456.2794Y} {456, 2794}

\bibitem[\protect\citeauthoryear{{Yang}, {Li}, {Looney}, {Girart}  \&
  {Stephens}}{{Yang} et~al.}{2017}]{Yang2017_nearfar}
{Yang} H.,  {Li} Z.-Y.,  {Looney} L.~W.,  {Girart} J.~M.,   {Stephens} I.~W.,
  2017, \mn@doi [\mnras] {10.1093/mnras/stx1951}, \href
  {http://adsabs.harvard.edu/abs/2017MNRAS.472..373Y} {472, 373}

\bibitem[\protect\citeauthoryear{{Yang}, {Li}, {Stephens}, {Kataoka}  \&
  {Looney}}{{Yang} et~al.}{2019}]{Yang2019}
{Yang} H.,  {Li} Z.-Y.,  {Stephens} I.~W.,  {Kataoka} A.,   {Looney} L.,  2019,
  \mn@doi [\mnras] {10.1093/mnras/sty3263}, \href
  {http://adsabs.harvard.edu/abs/2019MNRAS.483.2371Y} {483, 2371}

\bibitem[\protect\citeauthoryear{{Yen}, {Koch}, {Takakuwa}, {Krasnopolsky},
  {Ohashi}  \& {Aso}}{{Yen} et~al.}{2017}]{Yen2017}
{Yen} H.-W.,  {Koch} P.~M.,  {Takakuwa} S.,  {Krasnopolsky} R.,  {Ohashi} N.,
  {Aso} Y.,  2017, \mn@doi [\apj] {10.3847/1538-4357/834/2/178}, \href
  {https://ui.adsabs.harvard.edu/abs/2017ApJ...834..178Y} {834, 178}

\bibitem[\protect\citeauthoryear{{Zapata}, {Loinard}, {Rodr{\'\i}guez},
  {Hern{\'a}ndez-Hern{\'a}ndez}, {Takahashi}, {Trejo}  \& {Parise}}{{Zapata}
  et~al.}{2013}]{Zapata2013}
{Zapata} L.~A.,  {Loinard} L.,  {Rodr{\'\i}guez} L.~F.,
  {Hern{\'a}ndez-Hern{\'a}ndez} V.,  {Takahashi} S.,  {Trejo} A.,   {Parise}
  B.,  2013, \mn@doi [\apjl] {10.1088/2041-8205/764/1/L14}, \href
  {https://ui.adsabs.harvard.edu/abs/2013ApJ...764L..14Z} {764, L14}

\bibitem[\protect\citeauthoryear{{Zhu} et~al.,}{{Zhu} et~al.}{2019}]{Zhu2019}
{Zhu} Z.,  et~al., 2019, \mn@doi [\apjl] {10.3847/2041-8213/ab1f8c}, \href
  {https://ui.adsabs.harvard.edu/abs/2019ApJ...877L..18Z} {877, L18}

\bibitem[\protect\citeauthoryear{{van't Hoff}, {Tobin}, {Harsono}  \& {van
  Dishoeck}}{{van't Hoff} et~al.}{2018}]{vantHoff2018}
{van't Hoff} M. L.~R.,  {Tobin} J.~J.,  {Harsono} D.,   {van Dishoeck} E.~F.,
  2018, \mn@doi [\aap] {10.1051/0004-6361/201732313}, \href
  {https://ui.adsabs.harvard.edu/abs/2018A&A...615A..83V} {615, A83}

\makeatother
\end{thebibliography}

\bsp	
\label{lastpage}
\end{document}